\documentclass[aps,prb,showpacs,superscriptaddress]{revtex4}
\usepackage{latexsym}
\usepackage{amssymb}
\usepackage{graphicx}
\begin{document}
\title{Anomalous Hall effect in Rashba
two-dimensional electron systems based on narrow-band
semiconductors: side-jump and skew scattering mechanisms}
\author{S. Y. Liu}
\email{liusy@mail.sjtu.edu.cn}
\affiliation{Department of Physics and Engineering Physics, Stevens
Institute of Technology, Hoboken, New Jersey 07030, USA}
\affiliation{Department of Physics, Shanghai Jiaotong University, 1954
Huashan Road, Shanghai 200030, China}
\author{ Norman J. M. Horing}
\affiliation{Department of Physics and Engineering Physics, Stevens
Institute of Technology, Hoboken, New Jersey 07030, USA}
\author{X. L. Lei}
\affiliation{Department of Physics, Shanghai Jiaotong University, 1954
Huashan Road, Shanghai 200030, China}
\begin{abstract}
We employ a helicity-basis kinetic equation approach to investigate
the anomalous Hall effect in two-dimensional narrow-band
semiconductors considering both Rashba and extrinsic spin-orbit (SO)
couplings, as well as a SO coupling directly induced by an external
driving electric field. Taking account of long-range
electron-impurity scattering up to the second Born approximation, we
find that the various components of the anomalous Hall current fit
into two classes: (a) side-jump and (b) skew scattering anomalous
Hall currents. The side-jump anomalous Hall current involves
contributions not only from the extrinsic SO coupling but also from
the SO coupling due to the driving electric field. It also contains
a component which arises from the Rashba SO coupling and relates to
the off-diagonal elements of the helicity-basis distribution
function. The skew scattering anomalous Hall effect arises from the
anisotropy of the diagonal elements of the distribution function and
it is a result of both the Rashba and extrinsic SO interactions.
Further, we perform a numerical calculation to study the anomalous
Hall effect in a typical InSb/AlInSb quantum well. The dependencies
of the side-jump and skew scattering anomalous Hall conductivities
on magnetization and on the Rashba SO coupling constant are
examined.

\end{abstract}

\pacs{ 73.50.Dn, 72.20.Fr, 73.63.Hs}

\maketitle

\section{Introduction}
A nonvanishing magnetization in spin-split systems may lead to an
extraordinary Hall current.\cite{AHE} This so-called anomalous
Hall effect (AHE) was first observed more than a century
ago,\cite{Hall} but its complete understanding still remains a
challenge today.\cite{Review} Up to now, it has been made clear
that the AHE arises essentially from spin-orbit (SO) interactions,
which are the results of rapid movements of carriers in various
electric fields, such as nuclear fields ({\it e.g.} the
Dresselhaus SO coupling), electric fields associated with strains
or gate biases ({\it e.g.} the Rashba SO coupling), or fields
induced by electron-impurity scattering (extrinsic SO coupling),
{\it etc}.\cite{SO}

In 1954, Karplus and Luttinger proposed, for the first time, a
mechanism of the anomalous Hall effect.\cite{KL} This mechanism is
associated with the spin-orbit interaction due to nuclear fields
and yields to an anomalous Hall current (AHC) independent of any
electron-impurity scattering. Recently, it has been reformulated
by Jungwirth {\it et al.} within a framework of momentum-space
Berry phase,\cite{MacDonald1} and was used to explain the AHE in
various ferromagnetic systems, such as dilute magnetic
semiconductors (Ga,Mn)As,\cite{MacDonald1,MacDonald2}
ferromagnetic Fe,\cite{Yao} SrRuO series,\cite{SRO,SRO1,SRO2}
spinel CuCrSe {\it etc}.\cite{CCS}

The AHE may also stem from a spin-orbit coupling induced by
electron-impurity scattering, {\it i.e.} the extrinsic SO
coupling.\cite{Review} It was found that there are two mechanisms
responsible for this extrinsic AHE: a side-jump process proposed by
Berger\cite{Berger} and a skew scattering given by Smit.\cite{Smit}
The side-jump AHE arises from a sidewise shift of the center of the
electron wave packet and relates to an anomalous term in the current
operator caused by the extrinsic SO coupling.\cite{Berger,
Sinitsyn1} The skew scattering AHE corresponds to an anisotropic
enhancement of the wave packet due to electron-impurity scattering
and can be accounted for by considering the scattering in the second
Born approximation.\cite{Hugon} It was also clear that the side-jump
anomalous Hall conductivity is independent of impurity density
$n_i$, while the skew-scattering one is proportional to
$(n_i)^{-1}$. Recently, considering a short-range electron-impurity
scattering, Cr\'epieux, {\it et al.} presented a unified derivation
of both the side-jump and skew scattering mechanisms within the
framework of a formal Dirac equation for the electrons.\cite{Bruno1}
The weak-localization corrections to these anomalous Hall currents
have also been investigated.\cite{Bruno2,Bruno3,Wolfle1,Wolfle2}

The Rashba SO coupling in a two-dimensional (2D) electron system
with magnetization can also give rise to a nonvanishing contribution
to the Hall conductivity.\cite{Sinitsyn,Dugaev,Liu} It was found
that the anomalous Hall current due to the Rashba SO interaction
consists of two terms: one of which is associated with all electron
states below the Fermi surface and is independent of any
electron-impurity scattering; and another one relating only to
electrons near the Fermi surface, which is disorder-related but
independent of impurity density.\cite{Dugaev,Liu} Liu and Lei have
also clarified that, in the helicity basis, these two different
contributions to anomalous Hall current in 2D Rashba semiconductors
essentially arise from two distinct interband polarizations and they
relate to two distinct parts of the off-diagonal elements of the
distribution function.\cite{Liu} In these studies, the Rashba SO
coupling was considered nonperturbatively, but the extrinsic SO
interaction was completely ignored. Such a treatment is valid only
for 2D electron systems based on wide-band semiconductors. In 2D
narrow-band semiconductors, such as InSb/AlInSb quantum wells (QWs),
the coupling constant for extrinsic SO interaction is relatively
large (for example, the extrinsic SO coupling constant, $\lambda$,
is $\lambda=5.31$\,nm$^2$ for InSb, while it is equal to
$0.053$\,nm$^2$ for GaAs\cite{Halperin}). Hence, to investigate the
AHE in Rashba 2D systems based on narrow-band semiconductors, we
have to consider not only the Rashba but also the extrinsic SO
interactions.

In this paper, we employ a kinetic equation approach to investigate
the AHE in Rashba 2D narrow-band semiconductors. We deal with the
Rashba SO interaction in a nonperturbative way, while the extrinsic
SO coupling is considered in the first order of the coupling
constant $\lambda$. We also take account of the SO interaction
induced directly by the external driving electric field, which, to
our knowledge, was mentioned only by Nozi\`eres and Lewiner in the
absence of Rashba SO coupling.\cite{NL} In our study, to investigate
the skew-scattering AHE effect, we consider the electron-impurity
scattering up to the second-Born approximation. It is found that
various components of the anomalous Hall current can fit into two
classes: the side-jump and skew scattering anomalous Hall currents.
The side-jump AHC involves contributions from the extrinsic SO
coupling and SO coupling induced by the driving electric field. It
also contains a component which comes from the Rashba SO interaction
and relates to the off-diagonal elements of the helicity-basis
distribution function. The skew scattering AHC is associated with
anisotropic diagonal components of the distribution function and
stems from both the Rashba and extrinsic spin-orbit interactions. A
numerical calculation of the anomalous Hall current in a InSb/AlInSb
quantum well indicates that both the side-jump and skew scattering
anomalous Hall currents are of the same order of magnitude, leading
to complicated dependencies of the total anomalous Hall conductivity
on magnetization and on the Rashba spin-orbit coupling constant. It
is also clear that in the side-jump anomalous Hall current, the
contribution from SO coupling due to the driving electric field is
dominant for small magnetization.

This paper is organized as follows. In Sec. II, we derive the
kinetic equation for the nonequilibrium distribution function
considering long-range electron-impurity scattering up to the
second Born approximation. The solution of this equation to first
order of the extrinsic spin-orbit coupling is presented. We also
discuss the various components of the side-jump and skew
scattering anomalous Hall currents. In Sec. III, we perform a
numerical calculation to investigate the anomalous Hall effect in
a InSb/AlInSb quantum well. Finally, we review our results in Sec.
IV. The detailed form of the scattering term of the kinetic
equation is presented in an Appendix.

\section{Formalism}
\subsection{Hamiltonian and current operator}
We consider a Rashba quasi-two-dimensional narrow-band semiconductor
in the $x-y$ plane. When a homogeneous magnetization ${\bf
M}\equiv(0,0,M)$ induced by a weak magnetic field $B$, $M=g\mu_B B$
($g$ is the effective g-factor and $\mu_B$ is the Bohr magneton),
and a uniform in-plane dc electric field ${\bf E}$ are present, the
Hamiltonian of an electron with momentum ${\bf p}\equiv (p_x,p_y)
\equiv (p\cos \phi_{\bf p},p\sin \phi_{\bf p})$ can be written as
\begin{equation}
\check H=\check H_0 +\check H_{\rm imp}+\check H_E.\label{HH}
\end{equation}
${\check H}_0$ is the free electron Hamiltonian given by
\begin{equation}
{\check H}_0=\varepsilon_{p}+\alpha (\sigma^xp_y-\sigma^y
p_x)-M\sigma^z,\label{H0}
\end{equation}
where $\alpha$ is the Rashba SO coupling constant, $\sigma^l$
($l=x,y,z$) are the Pauli matrices, and $\varepsilon_p=p^2/2m^*$
with $m^*$ as the electron effective mass. By a local unitary spinor
transformation, $U_{\bf p}$,
\begin{equation}
U_{\bf p} =\frac{1}{\sqrt{2\lambda_p}}\left (
\begin{array}{cc}
\sqrt{\lambda_p+ M}&\sqrt{\lambda_p- M}\\
i{\rm e}^{i\phi_{\bf p}}\sqrt{ \lambda_p-M} &-i{\rm e}^{i\phi_{\bf
p}}\sqrt{ \lambda_p+M}
\end{array} \right ),
\end{equation}
Hamiltonian (\ref{H0}) can be diagonalized as  $\hat H_0\equiv
U_{\bf p}^+{\check H}_0U_{\bf p} ={\rm
diag}(\varepsilon_1(p),\varepsilon_2(p))$ with $\varepsilon_\mu
(p)=p^2/2m^*+(-1)^\mu \lambda_p$ ($\mu=1,2$ and $\lambda_p\equiv
\sqrt{M^2+\alpha^2 p^2}$) as dispersion relations of two
spin-orbit-coupled bands.

Since the extrinsic spin-orbit coupling in narrow-band
semiconductors cannot be ignored, the Hamiltonian $\check H_{\rm
imp}$, which describes the electron-impurity interaction, should
contain not only an ordinary scattering potential term but also a
term related to the extrinsic SO coupling:
\begin{equation}
\check H_{\rm imp}= \sum_i\left \{V(|{\bf r}-{\bf
R}_i|)-{\lambda}[{\bf \sigma}\times{\bf \nabla}V(|{\bf r}-{\bf
R}_i|)]\cdot {\bf p}\right\},
\end{equation}
where ${\bf r}$ and ${\bf R}_i$, respectively, are the coordinates
of the electron and impurity, $V(r)$ is the electron-impurity
scattering potential, and $\lambda$ is a spin-orbit coupling
constant depending on the intrinsic semiconductor parameters, such
as energy gap $E_0$, spin-orbit splitting $\Delta_{\rm SO}$, and
matrix element of the momentum operator between the conduction and
valence bands $P$: $\lambda=[1/E_0^2-1/(E_0+\Delta_{\rm
SO})^2]P^2/3$.\cite{Winkler} $H_E$ describes the application of the
external electric field, and, in the Coulomb gauge, it can be
written as
\begin{equation}
\check H_{E}=-e{\bf E}\cdot {\bf r}-{\lambda}[{\bf \sigma} \times
{\bf E} ]\cdot {\bf p}.\label{HE}
\end{equation}
In Eq.\,(\ref{HE}), we have considered the effect of the spin-orbit
coupling directly induced by the external driving dc electric field.

From Hamiltonian (\ref{HH}), it follows that, in spin basis, the
single-particle current operator, $\check {\bf j}({\bf p})$, can
be written as
\begin{equation}
\check j_l({\bf p})=\check j^{\rm f}_{l}({\bf p})+\check j^{\rm
imp}_{l}({\bf p}) +\check j^{ E}_{l}({\bf p}),
\end{equation}
with $l=x,y$. The term $\check j^{\rm f}_{l}({\bf p})$ comes from
the free-electron Hamiltonian $\check H_0$: $\check j^{\rm
f}_{l}({\bf p})=ep_l/m^*-\alpha\epsilon_{lmz}\sigma^m$ ($m=x,y$ and
$\epsilon_{lmz}$ is the totally antisymmetric tensor), while $\check
j^{\rm imp}_{l}({\bf p})$ comes from the SO coupling term of $\check
H_{\rm imp}$ and takes the form ($n=x,y$)
\begin{equation}
\check j^{\rm imp}_{l}({\bf p})={i\lambda e} \sum_{{\bf k},i}V_{{\bf
p}-{\bf k}}{\rm e}^{i{\bf R}_i\cdot({\bf k}-{\bf
p})}[\epsilon_{lmn}(k_m-p_m)\sigma^n].
\end{equation}
The term $\check j_l^{ E}({\bf p})$ arises from the spin-orbit
coupling directly induced by the external driving electric field and
it is given by
\begin{equation}
\check j_l^{ E}({\bf p})={-\lambda e^2}\epsilon_{lmn}\sigma^m E_n.
\end{equation}

Taking the statistical ensemble average, we find that the observed
net current, ${\bf J}$, consists of three components:
\begin{equation}
J_l=J_l^{\rm f}+J_l^{\rm imp}+J_l^{ E}.\label{TJ}
\end{equation}
$J_l^{{\rm f},E}$ is determined by $J_l^{{\rm f},E}=\sum_{\bf p}
{\rm Tr}[{\check j}_l^{{\rm f},E}({\bf p})\check \rho ({\bf p})]$,
with $\check \rho ({\bf p})$ as the distribution function in the
spin basis: $\check \rho_{\mu\nu} ({\bf p})=<\check\psi^+_{\nu\bf
p}\check \psi_{\mu\bf p}>$ ($\check \psi^+_{\mu\bf p}$ and $\check
\psi_{\mu\bf p}$, respectively, are the spin-basis electron creation
and annihilation operators). $J_l^{\rm imp}$ arises from the current
operator term $j_l^{\rm imp}({\bf p})$ and takes the form
\begin{equation}
J_l^{\rm imp}={i\lambda e}\sum_{{\bf p},{\bf k},i,\mu,\nu}V_{{\bf
p}-{\bf k}} {\rm e}^{i{\bf R}_i\cdot({\bf k}-{\bf
p})}\left\{<\check\psi^+_{\nu\bf p}\check \psi_{\mu\bf
k}>[\epsilon_{lmn}(k_m-p_m)\sigma_{\nu\mu}^n]\right\}.\label{JIMP}
\end{equation}
Obviously, to determine $J_l^{\rm imp}$, one has to analyze the
function $<\check\psi^+_{\nu\bf p}\check \psi_{\mu\bf k}>$.

Without loss of generality, we study here the anomalous Hall current
flow along the $x$-axis when the electric field is applied along the
$y$-direction, {\it i.e.} ${\bf E}=(0,E,0)$. In helicity basis, the
current $J_x^{ E}$ can be written as
\begin{equation}
J_x^{ E}={\lambda e^2E}\sum_{{\bf p}}[\hat \rho_{11}({\bf p})-\hat
\rho_{22}({\bf p})],\label{JE}
\end{equation}
with $\hat \rho_{\mu\nu}({\bf p})$ ($\mu,\nu=1,2$) defined as the
elements of the helicity-basis distribution function related to
the spin-basis distribution function by $\hat \rho({\bf
p})=U^+_{\bf p}\check \rho({\bf p})U_{\bf p}$. $J_x^{\rm f}$ can
be expressed as a sum of the contributions from the diagonal and
off-diagonal elements of the helicity-basis distribution function,
$J_x^{\rm fd}$ and $J_x^{\rm fo}$:
\begin{equation}
J_x^{\rm f}=J_x^{\rm fd}+J_x^{\rm fo},
\end{equation}
with
\begin{equation}
    J_x^{\rm fd}=e\sum_{\bf p}\left \{\left
    (\frac{1}{m^*}-\frac{\alpha^2}{\lambda_p}\right )p\cos\phi_{\bf p}\hat \rho_{11}({\bf p})
    +\left
    (\frac{1}{m^*}+\frac{\alpha^2}{\lambda_p}\right )
    p\cos\phi_{\bf p}\hat \rho_{22}({\bf p}) \right\},\label{DAHE}
\end{equation}
and
\begin{equation}
    J_x^{\rm fo}=e\sum_{\bf p}\left \{
   \frac{2\alpha M}{\lambda_p}\cos\phi_{\bf p}{\rm Re}
    \hat \rho_{12}({\bf p})+2\alpha\sin\phi_{\bf p}{\rm Im}
    \hat \rho_{12}({\bf p}) \right\}.\label{OAHE}
\end{equation}
From Eq.\,(\ref{OAHE}), we can see that the nonvanishing
contribution to $J_x^{\rm fo}$ comes from the component of the real
part of $\hat \rho_{12}({\bf p})$, depending on the momentum angle
through $\cos\phi_{\bf p}$: ${\rm Re}[(\hat \rho)_{12}({\bf
p})]=\xi(p)\cos\phi_{\bf p}+...$, and from the component of the
imaginary part of $\hat\rho_{12}({\bf p})$, involving $\sin\phi_{\bf
p}$: ${\rm Im}[(\hat \rho)_{12}({\bf p})]=\zeta (p)\sin\phi_{\bf
p}+...$. As a result, $J_x^{\rm fo}$ can be rewritten as
\begin{equation}
   J_x^{\rm fo}=2\alpha e^2E\sum_{{\bf p}}\left [\frac{
M\xi(p)}{\lambda_p}\cos^2\phi_{\bf p}+
    \zeta(p)\sin^2\phi_{\bf p}\right ].\label{OAHE1}
\end{equation}

Note that these $J_x$ components, $J_x^{\rm f}$, $J_x^{\rm imp}$,
and $J_x^{E}$, can fit into two classes: (a) side-jump and (b)
skew-scattering anomalous Hall currents. We know that the
side-jump AHE originates from the shift of the electron
wave-packet center towards the direction transverse to the driving
electric field. Such a shift is reflected by those current
operator components involving the antisymmetric tensor
$\epsilon_{lmn}$: $\check {j}_l^{\rm imp}$, $\check {j}_l^{E}$ and
the term in $\check {j}_l^{\rm f}$ associated with the Rashba
spin-orbit coupling. Correspondingly, the observed side-jump
anomalous Hall current $J_x^{\rm sj}$ is a sum of $J_x^{\rm imp}$,
$J_x^{ E}$ and $J_x^{\rm fo}$: $J_x^{\rm sj}=J_x^{\rm fo}+J_x^{\rm
imp}+J_x^{ E}$. From Eq.\,(\ref{DAHE}), we see that the remaining
$J_x$ component, $J_x^{\rm fd}$, becomes nonvanishing if there
exists a component of $\hat \rho_{\mu\mu}({\bf p})$ depending on
the angle of momentum through $\cos \phi_{\bf p}$. This implies
that $J_x^{\rm fd}$ results from an anisotropy due to
electron-impurity scattering and hence it just is a component of
$J_x^{\rm ss}$: $J_x^{\rm ss}=J_x^{\rm fd}$.

\subsection{Kinetic equation and its solution}

Obviously, in order to carry out the calculation of the anomalous
Hall current, it is necessary to determine the electron
distribution function. Under homogeneous and steady-state
conditions (averaging over a uniform impurity distribution), the
helicity-basis distribution function, $\hat\rho({\bf p})$, obeys a
kinetic equation written in the form
\begin{equation}
e{\bf E}\cdot \nabla_{\bf p}\hat\rho({\bf p})-e{\bf E}\cdot
[\hat\rho({\bf p}), U_{\bf p}^+{\bf \nabla}_{\bf p} U_{\bf p}]-{i
\lambda e}{\bf E}\cdot({\bf p}\times {\bf n})[U^+_{\bf
p}\sigma^zU_{\bf p},{\hat \rho}({\bf p})] +i[H_0,\hat\rho({\bf
p})]=-{\hat I},\label{KE}
\end{equation}
where ${\bf n}$ is a unit vector along $z$-axis and ${\hat I}$ is
a scattering term determined by
\begin{equation}
{\hat I}= \int \frac{d\omega}{2\pi}[{\hat \Sigma}^r({\bf
p},\omega){\hat {\rm G}}^<({\bf p},\omega)+{\hat \Sigma}^<({\bf
p},\omega){\hat {\rm G}}^a({\bf p},\omega)- {\hat {\rm G}}^r({\bf
p},\omega) {\hat \Sigma}^<({\bf p},\omega)-{\hat {\rm G}}^<({\bf
p},\omega){\hat \Sigma}^a({\bf p},\omega)].\label{CT}
\end{equation}
${\hat {\rm G}}^{r,a,<}({\bf p},\omega)$ and ${\hat
\Sigma}^{r,a,<}({\bf p},\omega)$ are, respectively, the
helicity-basis nonequilibrium Green's functions and self-energies.
Eq.\,(\ref{KE}) is derived from the Dyson equation of the
spin-basis nonequilibrium lesser Green's function by applying the
unitary transformation $U_{\bf p}$.

In Eq.\,(\ref{KE}), the electron-impurity scattering is embedded
in the self-energies, $\Sigma^{r,a,<}({\bf p},\omega)$. This
interaction in the helicity basis is described by a potential,
$\hat {\cal V}_{{\bf p}{\bf k}}$, which can be written as
\begin{equation}
\hat {\cal V}_{{\bf p}{\bf k}}=U_{\bf p}^+V({\bf p-k})U_{\bf
k}+{i\lambda}U_{\bf p}^+[{\bf n}\cdot ({\bf p}\times {\bf k})]V({\bf
p-k})U_{\bf k}.\label{V}
\end{equation}
In terms of the Feynman diagrams, $\hat {\cal V}_{{\bf p}{\bf k}}$
is denoted by two different interaction vertices: ordinary and
anomalous vertices, which, respectively, are depicted in
Figs.\,1(a) and 1(b). Since electron-impurity scattering will be
considered up to the second Born approximation, it is convenient
to express the self-energies by means of generalized T-matrices,
$\hat T^{r,a}_{\bf pk}(\omega)$, which obey the
equations\cite{Mahan}
\begin{equation}
\hat T^{r,a}_{\bf pk}(\omega)=\hat {\cal V}_{\bf pk}+\sum_{\bf
q}\hat {\cal V}_{\bf pq}{\hat {\rm G}}^{r,a}({\bf q},\omega)\hat
T^{r,a}_{\bf qk}(\omega).
\end{equation}
[These equations are exhibited in terms of Feynman diagrams in
Fig.\,1(c).] Thus, the lesser self-energy can be written
as\cite{Mahan}
\begin{equation}
\hat \Sigma^{<}({\bf p},\omega)=n_i \sum_{\bf k}\hat T^{r}_{\bf
pk}(\omega){\hat {\rm G}}^<({\bf k},\omega)\hat T^a_{\bf
kp}(\omega),
\end{equation}
while the retarded and advanced self-energies take the forms
\begin{equation}
\hat \Sigma^{r,a}({\bf p},\omega)=n_i \hat T^{r,a}_{\bf
pp}(\omega).
\end{equation}

In present paper, we restrict our considerations to the linear
response regime. In connection with this, all the functions, such
as the nonequilibrium Green's functions, self-energies, and
distribution function, can be expressed as sums of two terms: $A =
A_0 + A_1$, with $A$ representing the Green's functions,
self-energies or distribution function. $A_0$ and $A_1$,
respectively, are the unperturbed part and the linear electric
field part of $A$. In this way, the kinetic equation for the
linear electric field part of the distribution, $\hat \rho_1({\bf
p})$, can be written as
\begin{equation}
e{\bf E}\cdot \nabla_{\bf p}\hat\rho_0({\bf p})-e{\bf E}\cdot
[\hat\rho_0({\bf p}), U_{\bf p}^+{\bf \nabla}_{\bf p} U_{\bf
p}]-{i\lambda e}{\bf E}\cdot({\bf p}\times {\bf n})[U^+_{\bf
p}\sigma^zU_{\bf p},{\hat \rho}_0({\bf p})] +i[\hat
H_0,\hat\rho_1({\bf p})]=-{\hat I}^{(1)},\label{KE1}
\end{equation}
with $\hat I^{(1)}$ as the linear electric field part of the
collision term $\hat I$:
\begin{equation}
\hat I^{(1)}= \int \frac{d\omega}{2\pi}\left [{\hat
\Sigma}^<_{1}({\bf p},\omega){\hat {\rm G}}^a_{0}({\bf
p},\omega)-{\hat {\rm G}}^<_{1}({\bf p},\omega){\hat
\Sigma}^a_{0}({\bf p},\omega) +{\hat \Sigma}^r_{0}({\bf
p},\omega){\hat {\rm G}}^<_{1}({\bf p},\omega)- {\hat {\rm
G}}^r_{0}({\bf p},\omega) {\hat \Sigma}^<_{1}({\bf p},\omega)\right
].\label{CT1}
\end{equation}
We note that, here, the effect of $\hat {\rm G}_1^{r,a}({\bf
p},\omega)$ on distribution function has been ignored because
these linear electric parts of the retarded and advanced Green's
functions lead to a collisional broadening effect on $\hat
\rho_1({\bf p})$, which plays a secondary role in transport
studies.

To further simplify Eq.\,(\ref{KE1}), we employ a two-band
generalized Kadanoff-Baym ansatz (GKBA).\cite{GKBA,GKBA1} This
ansatz, which expresses the lesser Green's function through the
Wigner distribution function, has been proven sufficiently
accurate to analyze transport and optical properties in
semiconductors.\cite{Jauho} To first order in the dc field
strength, the GKBA reads,
\begin{equation}
\hat {\rm G}^<_{1}({\bf p},\omega)=-\hat {\rm G}_{0}^r({\bf
p},\omega)\hat \rho_1({\bf p})+\hat \rho_1({\bf p})\hat {\rm
G}_{0}^a({\bf p},\omega),\label{GKBA1}
\end{equation}
where the retarded and advanced Green's functions in helicity
basis are diagonal matrices: $\hat {\rm G}_0^{r,a}({\bf
p},\omega)={\rm diag}[(\omega-\varepsilon_1(p)\pm
i\delta)^{-1},(\omega-\varepsilon_2(p)\pm i\delta)^{-1}]$. Note
that the helicity-basis equilibrium distribution is also diagonal,
$\hat \rho_0({\bf p})={\rm diag}[n_{\rm
F}(\varepsilon_1(p)),n_{\rm F}(\varepsilon_2(p))]$ with $n_{\rm
F}(\omega)$ as the Fermi function.

Since our studies are concerned with electron transport within the
diffusive regime, it is sufficient to study the lowest order of
the distribution function in the impurity-density expansion. From
the diagonal parts of Eq.\,(\ref{KE1}), we see that the leading
order of the diagonal $\hat \rho_1({\bf p})$ elements, $(\hat
\rho_1)_{\mu\mu}({\bf p})$, is proportional to $(n_i)^{-1}$. These
diagonal elements of the distribution function give rise to
off-diagonal elements of the scattering term $\hat I^{(1)}$, which
are independent of the impurity density. From the fact that the
left-hand side of the off-diagonal parts of Eq.\,(\ref{KE1})
involves the term $i[\hat H_0,\hat\rho_1({\bf p})]$ proportional
to the off-diagonal elements of the distribution function:
\begin{equation}
i[\hat H_0,\hat\rho_1({\bf p})]=-2i\lambda_p\left (
\begin{array}{cc}
0&(\hat \rho_1)_{12}({\bf p})\\
(\hat \rho_1)_{21}({\bf p})&0
\end{array}
\right ),
\end{equation}
it follows that the leading order of the off-diagonal elements of
$\hat \rho_1({\bf p})$ should be of order $(n_i)^0$, {\it i.e.}
independent of the impurity density. Note that the contributions
to $\hat I^{(1)}$ from such off-diagonal elements of $\hat
\rho_1({\bf p})$ are linear in the impurity density and hence can
be ignored, while the contributions to $\hat I^{(1)}$ from the
diagonal elements, $(\hat \rho_1)_{\mu\mu}({\bf p})$, are
independent of $n_i$ and become dominant. Thus, to the
lowest-order of the $n_i$-power expansion, $\hat I^{(1)}$
effectively involves only the diagonal elements of the
distribution function.

From Eq.\,(\ref{KE1}) we see that the driving force of the kinetic
equation can be classified into two classes: diagonal $e{\bf
E}\cdot\nabla_{\bf p}\hat\rho_0$, and off-diagonal $-e{\bf E}\cdot
[\hat \rho_0({\bf p}), U_{\bf p}^+\nabla_{\bf p} U_{\bf p}]$ and
$-{i \lambda e}{\bf E}\cdot({\bf p}\times {\bf n})[U^+_{\bf
p}\sigma^zU_{\bf p},{\hat \rho}_0({\bf p})]$. In connection with
this, we may formally split the kinetic equation into two equations
with $\hat \rho_1({\bf p})=\hat \rho_1^{I}({\bf p})+\hat
\rho_1^{II}({\bf p})$ as
\begin{equation}
e{\bf E}\cdot \nabla_{\bf p} \hat \rho_0({\bf p})+i[\hat H_0,\hat
\rho_1^I({\bf p})]=-\hat I^{(1)},\label{EQ1}
\end{equation}
\begin{equation}
   -e{\bf E}\cdot [\hat \rho_0({\bf p}), U_{\bf p}^+\nabla_{\bf p} U_{\bf p}]
   -{i \lambda e}{\bf E}\cdot({\bf p}\times {\bf
n})[U^+_{\bf p}\sigma^zU_{\bf p},{\hat \rho}_0({\bf p})]
    +i[\hat H_0,\hat \rho_1^{II}({\bf p})]=0.\label{EQ2}
\end{equation}
From Eq.\,(\ref{EQ2}) it is evident that $\hat \rho_1^{II}({\bf
p})$ has null diagonal elements. Since $\hat I^{(1)}$ depends only
on the diagonal elements of the distribution function, $\hat
\rho_1^I({\bf p})$ and $\hat \rho_1^{II}({\bf p})$ can be
approximately determined independently of one another by
Eqs.\,(\ref{EQ1}) and (\ref{EQ2}).

Substituting the explicit forms of $\hat {\rm G}^{r,a}_0({\bf
p},\omega)$ into Eq.\,(\ref{CT1}) and considering
Eq.\,(\ref{GKBA1}), the elements of the linear electric-field
scattering term, $\hat I^{(1)}$, can be written as
\begin{eqnarray}
\hat I^{(1)}_{\mu\mu}&=&-2{\rm Im}\left [(\hat T^r_{\bf
pp})_{\mu\mu}(\varepsilon_\mu(p))\right ](\hat
\rho^I_1)_{\mu\mu}({\bf p})\nonumber\\
&& -2\pi\sum_{\bf
k}\delta[\varepsilon_\mu(p)-\varepsilon_\mu(k)](\hat T^r_{\bf
pk})_{\mu\mu}(\varepsilon_\mu(k))(\hat T^a_{\bf
kp})_{\mu\mu}(\varepsilon_\mu(k))(\hat\rho^I_1)_{\mu\mu}({\bf
k})\nonumber\\
&&-2\pi\sum_{\bf
k}\delta[\varepsilon_{\mu}(p)-\varepsilon_{\bar\mu}(k)](\hat
T^r_{\bf pk})_{\mu\bar \mu}(\varepsilon_{\bar \mu}(k))(\hat
T^a_{\bf kp})_{\bar\mu\mu}(\varepsilon_{\bar
\mu}(k))(\hat\rho^I_1)_{\bar\mu\bar\mu}({\bf k}),\label{I}
\end{eqnarray}
and
\begin{eqnarray}
\hat I^{(1)}_{\mu\bar\mu}&=&i(\hat T^r_{\bf
pp})_{\mu\bar\mu}(\varepsilon_{\bar\mu}(p))(\hat
\rho^I_1)_{\bar\mu\bar\mu}({\bf p})-i(\hat T^a_{\bf
pp})_{\mu\bar\mu}(\varepsilon_\mu(p))(\hat \rho^I_1)_{\mu\mu}({\bf
p})\label{II}
\\
&&+\sum_{{\bf k},\nu}\left \{(\hat T^r_{\bf
pk})_{\mu\nu}(\varepsilon_\nu(k))(\hat T^a_{\bf kp})_{\nu
\bar\mu}(\varepsilon_\nu(k))(\hat\rho^I_1)_{\nu\nu}({\bf k})
[(\hat{\rm G}^a_0)_{\bar\mu\bar\mu}({\bf
p},\varepsilon_\nu(k))-(\hat{\rm G}^r_0)_{\mu\mu}({\bf
p},\varepsilon_\nu(k))] \right \},\nonumber
\end{eqnarray}
with $\mu=1,2$ and $\bar \mu=3-\mu$.

Further, as in all previous studies, we consider the anomalous Hall
current only to the first order of the spin-orbit coupling constant
$\lambda$. Thus, the scattering term $\hat I^{(1)}$ and hence the
$T$-matrix may be considered only in the lowest- and first-order of
$\lambda$. On the other hand, we will evaluate the diagonal elements
of $\hat I^{(1)}$ up to the second-Born approximation, but its
off-diagonal elements only in the first-Born approximation. It is
widely accepted that, in usual cases, the self-consistent first Born
approximation is sufficiently accurate to analyze transport in
diffusive regime (correspondingly, the scattering term may be
considered only in the first-Born approximation). In our studies,
$\hat I^{(1)}_{\mu\mu}$ is evaluated up to the second Born
approximation because we should account for the skew scattering AHC
associated only with the diagonal elements of the distribution
function. Under these considerations, we need to analyze only the
Feynman diagrams depicted in Figs.\,1(a), 1(b), 1(d), and 1(e).
Substituting the $T$-matrix obtained from these diagrams, we obtain
the explicit form of the scattering term $\hat I^{(1)}$, which is
presented in the Appendix.

\begin{figure}
\includegraphics [width=0.45\textwidth,clip] {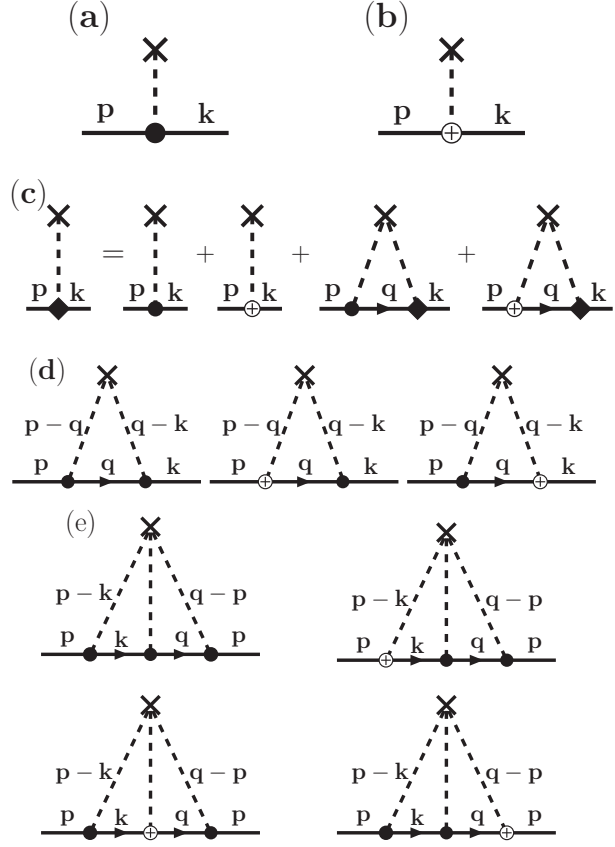}
\caption{Feynman diagrams for electron-impurity scattering. (a) and
(b) show the interaction vertices, which, respectively, correspond
to the original scattering potential and the potential due to the
extrinsic SO coupling. (c) is the equation for $T$-matrix. (d) and
(e), respectively, are Feynman diagrams for the $T$-matrix in the
first and second Born approximations up to the first order of the
spin-orbit coupling constant, $\lambda$.} \label{fig1}
\end{figure}

Considering the elastic features of the electron-impurity
scattering, Eq.\,(\ref{EQ1}) can be solved analytically. We know
that $\hat I^{(1)}$ does not involve the off-diagonal elements of
$\hat \rho_1^I({\bf p})$. Hence, the diagonal $\hat \rho_1^I({\bf
p})$ elements can be determined from the diagonal parts of
Eq.\,(\ref{EQ1}), while its off-diagonal elements are given by
substituting the obtained results for $(\hat \rho_1^I)_{\mu\mu}({\bf
p})$ into the off-diagonal parts of Eq.\,(\ref{EQ1}). We assume that
the solution $\hat \rho_1^I({\bf p})$ can be expressed as $\hat
\rho_1^I({\bf p})={\hat{\cal R}}_0({\bf p})+{\hat{\cal R}}_1({\bf
p})$ with ${\hat{\cal R}}_0({\bf p})$ and ${\hat{\cal R}}_1({\bf
p})$ as the lowest- and first-order terms of $\hat \rho_1^I({\bf
p})$ in the $\lambda$ expansion. In the lowest-order of $\lambda$,
the diagonal elements of the distribution function, ${\hat{\cal
R}}_{0\mu\mu}({\bf p})$, are determined by
\begin{equation}
e{\bf E}\cdot \nabla_{\bf p} \hat \rho_{0\mu\mu}({\bf p})=-{\cal
I}^{a}_{\mu}[{\hat{\cal R}}_{0}]-{\cal I}^{c}_{\mu}[{\hat{\cal
R}}_{0}],\label{LORD}
\end{equation}
where the diagonal terms of $\hat I^{(1)}$, ${\cal
I}^{a,c}_{\mu}[{\hat{\cal R}}_{0}]$, depend only on the diagonal
elements of ${\hat{\cal R}}_{0}({\bf p})$, ${\hat{\cal
R}}_{0\mu\mu}({\bf p})$, and are given by Eqs.\,(\ref{DDFs}) and
(\ref{Ic}). Since ${\cal I}^{c}_{\mu}[{\hat{\cal R}}_{0}]$ is a
higher-order term in the electron-impurity scattering, we can
assume that ${\cal I}^{c}_{\mu}[{\hat{\cal R}}_{0}]$ is much
smaller than ${\cal I}^{a}_{\mu}[{\hat{\cal R}}_{0}]$. Hence,
Eq.\,(\ref{LORD}) can be solved as follows: we first solve
Eq.\,(\ref{LORD}) by ignoring ${\cal I}^{c}_{\mu}[{\hat{\cal
R}}_{0}]$ and then substitute the obtained solution into ${\cal
I}^{c}_{\mu}$ to calculate a correction caused by ${\cal
I}^{c}_{\mu}$. Thus, we find that the solution of
Eq.\,(\ref{LORD}) consists of two terms: ${\hat{\cal
R}}_{0\mu\mu}({\bf p})={\hat{\cal R}}^{s}_{0\mu\mu}({\bf
p})+{\hat{\cal R}}^{c}_{0\mu\mu}({\bf p})$, with ${\hat{\cal
R}}^{s}_{0\mu\mu}({\bf p})$ and ${\hat{\cal R}}^{c}_{0\mu\mu}({\bf
p})$, respectively, determined by
\begin{equation}
e{\bf E}\cdot \nabla_{\bf p} \hat \rho_{0\mu\mu}({\bf p})=-{\cal
I}^{a}_{\mu}[{\hat{\cal R}}^{s}_{0}],\label{Rs}
\end{equation}
and
\begin{equation}
{\cal I}^{a}_{\mu}[{\hat{\cal R}}^{c}_{0}]+{\cal
I}^{c}_{\mu}[{\hat{\cal R}}^{s}_{0}]=0.\label{Rc}
\end{equation}
The solution of Eq.\,(\ref{Rs}), ${\hat{\cal
R}}^{s}_{0\mu\mu}({\bf p})$, depends on the momentum angle through
$\sin\phi_{\bf p}$ and takes the form, ${\hat{\cal
R}}^{s}_{0\mu\mu}({\bf p})=eE\Phi_{0\mu}^s(p)\sin\phi_{\bf p}$,
with the functions $\Phi^s_{0\mu}(p)$ given by
\begin{widetext}
\begin{equation}
    \Phi^s_{0\mu} (p)=-\frac{\partial n_{\rm F}(\varepsilon_\mu(p))}{\partial \varepsilon_\mu(p)}
        \frac{\left (\tau_{1{\bar \mu}{\bar \mu}}^{-1}
    +\tau_{2{\bar \mu} \mu}^{-1}\right )\frac{\partial \varepsilon_{\mu}(p)}{\partial p}
    +\tau_{3\mu{\bar \mu}}^{-1} \frac{\partial \varepsilon_{ \bar \mu}({\tilde p}_\mu)}
    { \partial {\tilde p}_\mu}}
    {\left (\tau_{1{\bar \mu}{\bar \mu}}^{-1}
    +\tau_{2{\bar \mu} \mu}^{-1}\right )
    \left (\tau_{1{\mu}{\mu}}^{-1}
    +\tau_{2\mu{\bar \mu}}^{-1}\right )-\tau_{3{\bar \mu}{\mu}}^{-1}
    \tau_{3\mu{\bar \mu}}^{-1}},\label{DDF1}
\end{equation}
\end{widetext}
where ${\tilde p}_\mu$ is given by equation $\varepsilon_{\bar
\mu}({\tilde p}_\mu)= \varepsilon_\mu (p)$ and the various
relaxation times $\tau_{i\mu\nu}$ ($i=1..3,\mu,\nu=1,2$) are
defined as
\begin{equation}
\frac {1}{\tau_{i\mu\nu}}=2\pi n_i\sum_{\bf k} |V({\bf p}-{\bf k})|^2
\Lambda_{i\mu\nu} ({\bf p},{\bf k}),
\end{equation}
with $\Lambda_{1\mu\nu}({\bf p},{\bf k})=\frac 12[1-\cos (\phi_{ \bf
p}-\phi_{\bf k})]a_+({\bf p},{\bf k}) \Delta_{\mu\nu}$,
$\Lambda_{2\mu\nu}({\bf p},{\bf k})=\frac 12 a_-({\bf p},{\bf k})
\Delta_{\mu\nu}$, $\Lambda_{3\mu\nu}({\bf p}-{\bf k})=\frac 12 \cos
(\phi_{ \bf p}-\phi_{\bf k}) a_-({\bf p},{\bf k}) \Delta_{\mu\nu}$
and $a_\pm({\bf p},{\bf k})\equiv (\lambda_p\lambda_k\pm M^2\pm
\alpha^2kp\cos (\phi_{\bf p}-\phi_{\bf k}))/ \lambda_p\lambda_k $.
Substituting the result for ${\hat{\cal R}}^{s}_{0\mu\mu}({\bf p})$
into ${\cal I}^{c}_{\mu}$, we find that the solution of
Eq.\,(\ref{Rc}), ${\hat{\cal R}}^{c}_{0\mu\mu}({\bf p})$, relates to
a cosine function of the momentum angle: ${\hat{\cal
R}}_{0\mu\mu}({\bf p})=eE\Phi_{0\mu}^c(p)\cos\phi_{\bf p}$.
$\Phi_{0\mu}^{c}(p)$ is given by
\begin{equation}
\Phi_{0\mu}^{c}(p)= - \frac{\left (\tau_{1{\bar \mu}{\bar
\mu}}^{-1}
    +\tau_{2{\bar \mu} \mu}^{-1}\right ){\cal L}_{0\mu}^{c}(p)
    +\tau_{3\mu{\bar \mu}}^{-1} {\cal L}_{0\bar\mu}^{c}(\tilde p_\mu)}
    {\left (\tau_{1{\bar \mu}{\bar \mu}}^{-1}
    +\tau_{2{\bar \mu} \mu}^{-1}\right )
    \left (\tau_{1{\mu}{\mu}}^{-1}
    +\tau_{2\mu{\bar \mu}}^{-1}\right )-\tau_{3{\bar \mu}{\mu}}^{-1}
    \tau_{3\mu{\bar \mu}}^{-1}},\label{phic}
\end{equation}
with
\begin{eqnarray}
{\cal L}_{0\mu}^{c}(p)&=& \pi^2 \alpha^2Mn_i\sum_{{\bf k},{\bf
q}}V_{{\bf p} -{\bf k}}V_{{\bf k} -{\bf q}}V_{{\bf q} -{\bf
p}}\frac{1}{\lambda_p\lambda_k}\sin(\phi_{\bf k}-\phi_{\bf p})
\nonumber\\
&&\times \{c_{\mu +}({\bf p},{\bf q},{\bf
k})\Delta_{\mu\mu}\Phi_{0\mu}^s( p)+c_{\mu -}({\bf p},{\bf q},{\bf
k})\Delta_{\mu\bar \mu}\Phi^s_{0\bar \mu}(\tilde p_\mu)\},
\end{eqnarray}
and $c_{\mu \pm}({\bf p},{\bf q},{\bf k})$ defined by
Eq.\,(\ref{c}).

The off-diagonal ${\hat{\cal R}}_{0}({\bf p})$ elements,
${\hat{\cal R}}_{0\mu\bar\mu}({\bf p})$, are obtained by
substituting ${\hat{\cal R}}_{0\mu\mu}^s({\bf p})$ into the
off-diagonal parts of Eq.\,(\ref{EQ1}):
\begin{equation}
i[\hat H_0,{\hat{\cal R}}_{0}({\bf p})]=-{\cal S}^{a}[{\hat{\cal
R}}_{0}^s].
\end{equation}
Here, the effect of ${\hat{\cal R}}_{0\mu\mu}^c({\bf p})$ on the
off-diagonal elements was ignored. We find that ${\hat{\cal
R}}_{0\mu\bar\mu}({\bf p})$ depends on the momentum angle not only
through $\sin \phi_{\bf p}$ but also through $\cos \phi_{\bf p}$.
However, as discussed above, we are interested only in the part of
${\rm Re}[({\hat{\cal R}}_0)_{12}({\bf p})]$, which depends on
$\phi_{\bf p}$ through $\cos\phi_{\bf p}$, and the part of ${\rm
Im}[({\hat{\cal R}}_0)_{12}({\bf p})]$, involving $\sin \phi_{\bf
p}$: ${\rm Re}[({\hat{\cal R}}_0)_{12}({\bf p})]=
\xi^I_0(p)\cos\phi_{\bf p}+...$ and ${\rm Im}[({\hat{\cal
R}}_0)_{12}({\bf p})]=\zeta^I_0(p)\sin\phi_{\bf p}+...$, with
$\xi^I_0(p)$ and $\zeta^I_0(p)$ taking the forms,
\begin{eqnarray}
    \zeta_0^I(p)&=&\frac{eE\pi n_i}{4\lambda_p}\sum_{{\bf k}\,\mu=1,2}|V({\bf
p}-{\bf k})|^2{\rm Im}[g_a({\bf p},{\bf k})](-1)^{\mu}
\{\Delta_{\mu\mu}\Phi^s_{0\mu}(p)[1-\cos (\phi_{ \bf p}-\phi_{\bf
k})]\nonumber\\
&&-\Delta_{\mu{\bar \mu}} [\Phi^s_{0\mu}(p)-\Phi^s_{0\bar
\mu}(\tilde p_\mu) \cos (\phi_{ \bf p}-\phi_{\bf k})]\},\label{z1}
\end{eqnarray}
and
\begin{eqnarray}
    \xi_0^I(p)&=&\frac{eE\pi n_i}{4\lambda_p}\sum_{{\bf k}\,\mu=1,2}|V({\bf
p}-{\bf k})|^2{\rm Re}[g_a({\bf p},{\bf k})](-1)^{\mu} \nonumber\\
&&\times\left \{\Delta_{\mu\mu}\Phi^s_{0\mu}(p)\sin (\phi_{ \bf
p}-\phi_{\bf k})
 -\Delta_{\mu{\bar \mu}} \Phi^s_{0\bar \mu}(\tilde p_\mu)
\sin (\phi_{ \bf p}-\phi_{\bf k})\right \},\label{x1}
\end{eqnarray}
and $g_{a}({\bf p},{\bf k})$ defined as $g_{a}({\bf p},{\bf
k})\equiv [\alpha k\lambda_{p}\sin (\phi_{\bf p}-\phi_{\bf k})
+i\alpha M(p- k \cos(\phi_{\bf p}-\phi_{\bf
k}))]/(\lambda_k\lambda_p)$.

 To obtain the first-order term of the distribution function in
 the $\lambda$ expansion, ${\hat{\cal R}}_{1}({\bf p})$, we
substitute ${\hat{\cal R}}_{0\mu\mu}^{s}({\bf p})$ into the
diagonal components of the scattering term, ${\cal I}^{b}_\mu$ and
${\cal I}^{d}_\mu$, as well as its off-diagonal component ${\cal
S}^{b}$. We find that ${\cal I}^{b}_\mu[{\hat{\cal R}}^s_{0}]$
depends on the angle of momentum through a sine function, while
${\cal I}^{d}_\mu[{\hat{\cal R}}^s_{0}]$ relates to a cosine
function: ${\cal I}^{b}_\mu[{\hat{\cal R}}^s_{0}]= {\cal
L}^{s}_{1\mu}(p)\sin\phi_{\bf p}$ and ${\cal
I}^{d}_{1\mu}[{\hat{\cal R}}^s_{0}]={\cal
L}^{c}_{1\mu}(p)\cos\phi_{\bf p}$ with ${\cal L}^{c,s}_{1\mu}$
given by
\begin{eqnarray}
{\cal L}_{1\mu}^{s}(p)&=& 2\pi n_i\lambda\alpha^2 \sum_{\bf k}
|V({\bf p} -{\bf k})|^2
\frac{k^2p^2}{\lambda_p\lambda_k}\sin^2(\phi_{\bf k}-\phi_{\bf
p}) \nonumber\\
&&\times\{[1-\cos(\phi_{\bf k}-\phi_{\bf p})] \Phi^s_{0\mu}(
p)\Delta_{\mu\mu}-[\Phi^s_{0\mu}(p)-\Phi^s_{0\bar \mu}({\tilde
p_\mu})\cos(\phi_{\bf k}-\phi_{\bf p})]\Delta_{\mu{\bar \mu}}\}
\end{eqnarray}
and
\begin{eqnarray}
{\cal L}_{1\mu}^{c}(p)&=& \pi^2 \lambda Mn_i\sum_{{\bf k},{\bf
q}}V_{{\bf p} -{\bf k}}V_{{\bf k} -{\bf q}}V_{{\bf q} -{\bf
p}}\frac{1}{\lambda_p\lambda_k}\sin(\phi_{\bf k}-\phi_{\bf
p})\nonumber\\
&&\times \{d_{\mu +}({\bf p},{\bf q},{\bf
k})\Delta_{\mu\mu}\Phi_{0\mu}^s( p)+d_{\mu -}({\bf p},{\bf q},{\bf
k})\Delta_{\mu\bar \mu}\Phi^s_{0\bar \mu}(\tilde p_\mu)\}.
\end{eqnarray}
Here, $d_{\mu \pm}({\bf p},{\bf q},{\bf k})$ are defined by
Eq.\,(\ref{d}). From the diagonal parts of Eq.\,(\ref{EQ1}) in the
first order of $\lambda$, ${\cal I}_\mu^{a}[{\hat{\cal R}}_1]+{\cal
I}_\mu^{b}[{\hat{\cal R}}^s_0]+{\cal I}_\mu^{d}[{\hat{\cal
R}}^s_0]=0$, it follows that ${\hat{\cal R}}_{1\mu\mu}({\bf p})$ can
be written as
\begin{equation}
{\hat{\cal R}}_{1\mu\mu}({\bf p})=eE\Phi_{1\mu}^s(p)\sin\phi_{\bf
p}+eE\Phi_{1\mu}^c(p)\cos\phi_{\bf p},\label{R}
\end{equation}
with $\Phi_{1\mu}^{s,c}(p)$ determined by
\begin{equation}
\Phi_{1\mu}^{s,c}(p)= - \frac{\left (\tau_{1{\bar \mu}{\bar
\mu}}^{-1}
    +\tau_{2{\bar \mu} \mu}^{-1}\right ){\cal L}_{1\mu}^{s,c}(p)
    +\tau_{3\mu{\bar \mu}}^{-1} {\cal L}_{1\bar\mu}^{s,c}(\tilde p_\mu)}
    {\left (\tau_{1{\bar \mu}{\bar \mu}}^{-1}
    +\tau_{2{\bar \mu} \mu}^{-1}\right )
    \left (\tau_{1{\mu}{\mu}}^{-1}
    +\tau_{2\mu{\bar \mu}}^{-1}\right )-\tau_{3{\bar \mu}{\mu}}^{-1}
    \tau_{3\mu{\bar \mu}}^{-1}}.\label{DDF2}
\end{equation}

The off-diagonal elements of ${\hat{\cal R}}_1({\bf p})$ are
obtained from the off-diagonal parts of Eq.\,(\ref{EQ1}) in the
first order of $\lambda$: $i[\hat H_0,{\hat{\cal R}}_{1}({\bf
p})]=-{\cal S}^{b}[{\hat{\cal R}}_{0}^s]$. According to the
definitions, ${\rm Re}[({\hat{\cal R}}_1)_{12}({\bf
p})]=\xi_1^I(p)\cos\phi_{\bf p}+...$ and ${\rm Im}[({\hat{\cal
R}}_1)_{12}({\bf p})]=\zeta_1^I(p)\sin\phi_{\bf p}+...$,
$\xi_{1}^I(p)$ and $\zeta_{1}^I(p)$ can be written as
\begin{eqnarray}
    \zeta_{1}^I(p)&=&\frac{eE\pi n_i}{4\lambda_p}\sum_{{\bf k}\,\mu=1,2}|V({\bf
p}-{\bf k})|^2{\rm Im}[g_{b}({\bf p},{\bf k})](-1)^{\mu}
\{\Delta_{\mu\mu}\Phi^s_{0\mu}(p)[1-\cos (\phi_{ \bf p}-\phi_{\bf
k})]\nonumber\\
&&-\Delta_{\mu{\bar \mu}} [\Phi^s_{0\mu}(p)-\Phi^s_{0\bar
\mu}(\tilde p_\mu) \cos (\phi_{ \bf p}-\phi_{\bf k})]\},\label{z1b}
\end{eqnarray}
and
\begin{equation}
    \xi_{1}^I(p)=\frac{eE\pi n_i}{4\lambda_p}\sum_{{\bf k}\,\mu=1,2}|V({\bf
p}-{\bf k})|^2{\rm Re}[g_{b}({\bf p},{\bf k})](-1)^{\mu} \sin
(\phi_{ \bf p}-\phi_{\bf k})\left
\{\Delta_{\mu\mu}\Phi^s_{0\mu}(p)-\Delta_{\mu{\bar \mu}}
\Phi^s_{0\bar \mu}(\tilde p_\mu) \right \}.\label{x1b}
\end{equation}

The other component of the kinetic equation, Eq.\,(\ref{EQ2}), can
be solved easily. The solution $\hat\rho_1^{II}({\bf p})$ has null
diagonal elements. Its off-diagonal elements can be written as
$(\rho_1^{II})_{12}({\bf p})=\xi^{II}(p)\cos\phi_{\bf
p}+i\zeta^{II}(p)\sin\phi_{\bf p}$ with $\xi^{II}(p)$ and
$\zeta^{II}(p)$ defined as
\begin{equation}
    \xi^{II}(p)=\frac{\alpha e E}{4\lambda_p^2}\left (
1 -2\lambda p^2\right )
    \{n_{\rm F}[\varepsilon_{1}(p)]-n_{\rm
    F}[\varepsilon_{2}(p)]\}\label{xi2}
\end{equation}
and
\begin{equation}
    \zeta^{II}(p)=\frac{\alpha e EM}{4\lambda_p^3}
    \{n_{\rm F}[\varepsilon_{1}(p)]-n_{\rm
    F}[\varepsilon_{2}(p)]\}.\label{ze2}
\end{equation}

\subsection{Anomalous Hall current}

We first analyze the component of the anomalous Hall current,
$J^{\rm f}_x$, that is associated with the current operator term
arising from the free-electron Hamiltonian and is a sum of
contributions from the diagonal and off-diagonal elements of the
distribution function: $J_x^{\rm f}=J^{\rm fd}_x+J^{\rm fo}_x$. We
know that $J_x^{\rm fd}$ is a component of the skew-scattering
AHC. Considering Eq.\,(\ref{DAHE}), it is obvious that the
nonvanishing $J^{\rm fd}_x$ comes from the diagonal terms of the
distribution function, ${\hat{\cal R}}^c_{0\mu\mu}({\bf p})$ and
${\hat{\cal R}}^c_{1\mu\mu}({\bf p})$, which depend on momentum
angle through the cosine function. Thus, $J_x^{\rm fd}$ can be
written as $J_x^{\rm fd}=J_x^{\rm ss-L}+J_x^{\rm ss-F}$, where
$J_x^{\rm ss-L}$ and $J_x^{\rm ss-F}$, respectively, are
associated with the lowest- and first-order terms of the
distribution function, ${\hat{\cal R}}^c_{0\mu\mu}({\bf p})$ and
${\hat{\cal R}}^c_{1\mu\mu}({\bf p})$:
\begin{equation}
J_x^{\rm ss-L}=e^2E\sum_{{\bf p},\mu}\left \{\left
    (\frac{1}{m^*}+(-1)^\mu\frac{\alpha^2}{\lambda_p}\right )p\cos^2\phi_{\bf p}\Phi^c_{0\mu}(p)
    \right \},
\end{equation}
\begin{equation}
J_x^{\rm ss-F}=e^2E\sum_{{\bf p},\mu}\left \{\left
    (\frac{1}{m^*}+(-1)^\mu\frac{\alpha^2}{\lambda_p}\right )p\cos^2\phi_{\bf p}\Phi^c_{1\mu}(p)
 \right \}.
\end{equation}
Note that both $J_x^{\rm ss-L}$ and $J_x^{\rm ss-F}$ are
proportional to the inverse of the impurity-density, {\it i.e.}
$(n_i)^{-1}$, appearing when electron-impurity scattering is
considered up to the second-Born approximation.

Since both the distribution terms $\hat {\rho}^{I}({\bf p})$ and
$\hat {\rho}^{II}({\bf p})$ have nonvanishing off-diagonal
elements, the contribution to the anomalous Hall current from
off-diagonal elements of $\hat {\rho}({\bf p})$, $J_x^{\rm fo}$,
can be expressed as $J_x^{\rm fo}=\left. J_x^{\rm
fo}\right|^{I}+\left. J_x^{\rm fo}\right|^{II}$, where $\left.
J_x^{\rm fo}\right|^I$ and $\left. J_x^{\rm fo}\right|^{II}$ arise
from $\hat {\rho}^{I}_{12}({\bf p})$ and $\hat
{\rho}^{II}_{12}({\bf p})$, respectively, and take the forms
\begin{eqnarray}
\left .J_{x}^{\rm fo}\right|^{I}&=&2e^2E\sum_{\bf p}\left
\{\frac{\alpha M}{\lambda_p}\left [\xi_0^I(p)+\xi_1^I(p)\right
]\cos^2\phi_{\bf p}+\alpha\left [\zeta_0^I(p)+\zeta_1^I(p)\right
]\sin^2\phi_{\bf p}\right\},\label{AHEfoI}
\end{eqnarray}
and
\begin{eqnarray}
\left .J_{x}^{\rm fo}\right|^{II}&=&2e^2E\sum_{\bf p}\left
\{\frac{\alpha M}{\lambda_p}\xi^{II}(p)\cos^2\phi_{\bf
p}+\alpha\zeta^{II}(p)\sin^2\phi_{\bf
p}\right\}\nonumber\\
&=& \frac{M\alpha^2 e^2E}{2}\sum_{\bf p}\frac{1}{\lambda_p^3} \left
(1-{\lambda p^2}\right )\{n_{\rm F}[\varepsilon_1(p)]-n_{\rm
F}[\varepsilon_2(p)]\}.\label{AHEfoII}
\end{eqnarray}

From Eq.\,(\ref{AHEfoI}) we see that $\left .J_{x}^{\rm
fo}\right|^{I}$ is independent of the impurity density, due to the
$n_i$-independence of $\xi^I_{0}$ and $\zeta^I_{1}$. However, it is
due to disorder and relates to longitudinal transport. It is obvious
that $\left. J_x^{\rm fo}\right|^I$ involves the derivative of the
equilibrium distribution function, {\it i.e.} $\partial n_{\rm
F}(\omega)/\partial \omega$. This implies that $\left .J_x^{\rm
fo}\right |^I$ arises only from electron states in the vicinity of
the Fermi surface, or, in other words, from electron states involved
in longitudinal transport. Physically, the electrons participating
in transport experience impurity scattering, producing diagonal
$\hat \rho_1^I({\bf p})$ elements of order of $n_i^{-1}$. Moreover,
the scattering of these perturbed electrons by impurities also gives
rise to an interband polarization, which eliminates dependence on
the impurity density within the diffusive regime.

However, the anomalous Hall current $\left .J_x^{\rm fo}\right
|^{II}$ is a function of the entire unperturbed equilibrium
distribution, $n_{\rm F}(\omega)$, not just of its derivative,
$\partial n_{\rm F}(\omega)/\partial \omega$, at the Fermi surface.
This indicates that $\left .J_x^{\rm fo}\right |^{II}$ has
contributions from all electron states below the Fermi sea.
Obviously, $\left .J_x^{\rm fo}\right |^{II}$ is independent of any
electron-impurity scattering and relates to the driving terms, one
of which is just the interband electric dipole moment, while the
other one arises from the SO coupling directly induced by the
driving electric field.

From Eq.\,(\ref{JIMP}), it is obvious that to determine the
first-order term of $J_x^{\rm imp}$ in the $\lambda$-power
expansion, one has to deal with the function $<\check\psi^+_{\nu\bf
p}\check \psi_{\mu\bf k}>$ in the lowest order of $\lambda$. We find
that this lowest-order term of $<\check\psi^+_{\nu\bf p}\check
\psi_{\mu\bf k}>$ can be evaluated from the kinetic equation for the
distribution function $\check \rho({\bf p})$. To show this, we start
with a Heisenberg equation for the operator $\check\psi^+_{\nu\bf
p}\check\psi_{\mu\bf p}$:
\begin{equation}
i\hbar \frac{\partial }{\partial T}\check\psi^+_{\nu\bf
p}\check\psi_{\mu\bf p}= [\check H,\check\psi^+_{\nu\bf
p}\check\psi_{\mu\bf p}]=-ie{\bf E}\cdot {\bf \nabla}_{\bf
p}[\check\psi^+_{\nu\bf p}\check\psi_{\mu\bf p}]+[{\check
H}_0+{\check H}_E,\check\psi^+_{\nu\bf p}\check\psi_{\mu\bf p}]
+\check I^s_{\mu\nu}({\bf p}),\label{KK}
\end{equation}
where, $\check I_{\mu\nu}^{ s}({\bf p})\equiv [\check H_{\rm
imp},\check\psi^+_{\nu\bf p}\check\psi_{\mu\bf p}]$. In the lowest
order of $\lambda$, $\check I^{s}_{\mu\nu}({\bf p})$ takes the form,
\begin{equation}
\check I^{s}_{\mu\nu}({\bf p})\approx \sum_{{\bf k},i}V_{{\bf
p}-{\bf k}}[{\rm e}^{i{\bf R}_i\cdot({{\bf p}-{\bf
k}})}\check\psi^+_{\nu\bf k}\check \psi_{\mu\bf p}-{\rm e}^{i{\bf
R}_i\cdot({{\bf k}-{\bf p}})}\check\psi^+_{\nu\bf p}\check
\psi_{\mu\bf k}].\label{Is}
\end{equation}
Multiplying both sides of Eq.\,(\ref{KK})  by
$\varepsilon_{lmn}p_m\sigma^n_{\nu\mu}$ and taking the summation
over $\mu$ and $\nu$, we get
\begin{equation}
\sum_{{\bf p},\mu\nu}\varepsilon_{lmn}[p_m\sigma^n_{\nu\mu} <\check
I^{s}_{\mu\nu}({\bf p})>]\approx \sum_{{\bf p},{\bf k},\mu\nu}{\rm
e}^{i{\bf R}_i\cdot({{\bf k}-{\bf p}})}V_{{\bf k}-{\bf
p}}\varepsilon_{lmn}(k_m-p_m)[<\check\psi^+_{\nu\bf
p}\check\psi_{\mu\bf k}>\sigma^n_{\nu\mu}],\label{Is2}
\end{equation}
with $<...>$ denoting a statistical average. Obviously, the
right-hand side of Eq.\,(\ref{Is2}) is just the AHC component
$J_l^{\rm imp}$. On the other hand, taking the statistical average
of Eq.\,(\ref{KK}) reduces it to the kinetic equation for $\check
\rho_{\mu\nu}({\bf p})=<\check\psi^+_{\nu\bf p}\check\psi_{\mu\bf
p}>$. Hence, from Eqs.\,(\ref{KK}) and (\ref{Is2}), it follows that
$J_l^{\rm imp}$ can be written in the spin basis as
\begin{equation}
J_l^{\rm imp}=-{\lambda e}\sum_{\bf p}\varepsilon_{lmn}p_m {\rm
Tr}\left \{\sigma^n \left (e{\bf E}\cdot {\bf \nabla}_{\bf p}\check
\rho({\bf p})+i[\check H_0,\check \rho({\bf p}) ]\right )\right \},
\end{equation}
where the contribution associated with $\check H_E$ is ignored
because it is of higher order in $\lambda$. $J_l^{\rm imp}$ can also
be determined in the helicity basis by means of
\begin{equation}
J_l^{\rm imp}=-{\lambda e}\sum_{\bf p}\varepsilon_{lmn}p_m {\rm
Tr}\left \{U_{\bf p}^+\sigma^n U_{\bf p}\left [e{\bf E}\cdot {\bf
\nabla}_{\bf p}\hat \rho({\bf p}) -e{\bf E}\cdot [\hat \rho({\bf
p}),U^+_{\bf p}{\bf \nabla}_{\bf p}U_{\bf p}]+i[\hat H_0,\hat
\rho({\bf p}) ]\right ]\right \}.\label{J1}
\end{equation}
In the linear response regime, Eq.\,(\ref{J1}) reduces to
\begin{eqnarray}
J_x^{\rm imp}&=&-{\lambda e}\sum_{\bf p}\frac{p_y}{\lambda_p} \left
\{eEM\nabla_{p_y}[(\hat \rho_0)_{11}({\bf p})-(\hat
\rho_0)_{22}({\bf p})]\right.
\nonumber\\
&&\left. -\frac{eE\alpha^2p}{\lambda_p^2}M\sin\phi_{\bf p} [(\hat
\rho_0)_{11}({\bf p})-(\hat \rho_0)_{22}({\bf p})]+4\alpha
p\lambda_p[\zeta_0^I(p)+\zeta^{II}(p)]\sin\phi_{\bf p}\right\}\nonumber\\
&=&-{\lambda e}\sum_{\bf p} \left \{-\frac{eEM}{\lambda_p}\left
\{n_F[\varepsilon_{1}({p})]-n_F[\varepsilon_2({ p})]\right
\}+4\alpha pp_y[\zeta_0^I(p)+\zeta^{II}(p)]\sin\phi_{\bf
p}\right\},\label{Jximp}
\end{eqnarray}
To obtain the last equality in Eq.\,(\ref{Jximp}), the momentum
integral with integrand involving the derivative with respect to
$p_y$ was performed by parts. Note that Sinitsyn, {\it et al.}
recently investigated this component of the anomalous Hall current
in the absence of Rashba SO coupling by analyzing the effect of
impurity scattering on the coordinate shift of the electron
wave-packet.\cite{Sinitsyn1}

$J_x^{E}$ arises from SO coupling directly induced by the driving
electric field. Considered to linear order in the electric field,
it takes the form
\begin{equation}
J_x^{E}={\lambda e^2E}\sum_{{\bf p}}\left
\{n_F[\varepsilon_{1}({p})]-n_F[\varepsilon_2({ p})]\right\}.
\end{equation}
Obviously, this contribution to the anomalous Hall current is
independent of any electron-impurity scattering. If only one of
the parameters-the Rashba SO coupling constant or the
magnetization-is zero and other remains finite, $J_x^{E}$ doesn't
vanish. In contrast to this, $J^{\rm fd}_x$ and $J^{\rm fo}_x$, as
well as $J^{\rm imp}_x$ reduce to zero for just one of them
vanishing, null $\alpha$ {\it or} $M$. Note that in all previous
studies, the contribution to anomalous Hall current from the SO
term due to the driving electric field has been ignored. In the
following numerical calculation, we will show that $J_x^{E}$ plays
an important role, especially in the low magnetization regime.

Thus, after all components of AHC are determined, the total
anomalous Hall current can be obtained from Eq.\,(\ref{TJ}). We
define the total anomalous Hall conductivity as
$\sigma_{xy}=J_x/E$. Obviously, $\sigma_{xy}$ can be written as
\begin{equation}
\sigma_{xy}=\sigma_{xy}^{\rm sj}+\sigma_{xy}^{\rm ss},
\end{equation}
with $\sigma_{xy}^{\rm sj}=J_x^{\rm sj}/E=\sigma_{xy}^{\rm fo}+\sigma_{xy}^{\rm
imp}+\sigma_{xy}^{E}$ and $\sigma_{xy}^{\rm ss}=J_x^{\rm ss}/E=\sigma_{xy}^{\rm ss-L}+
\sigma_{xy}^{\rm ss-F}$. Here, the quantities, $\sigma_{xy}^{\rm fo}$, $\sigma_{xy}^{\rm
imp}$, $\sigma_{xy}^{E}$, $\sigma_{xy}^{\rm ss-L}$, and
$\sigma_{xy}^{\rm ss-L}$, are defined as $\sigma_{xy}^{{\rm imp},E}=J_x^{{\rm
imp},E}/E$, $\sigma_{xy}^{\rm fo}=J_{x}^{\rm fo}/E$,
$\sigma_{xy}^{\rm
ss-L}= J_x^{\rm ss-L}/E$ and $\sigma_{xy}^{\rm ss-F}= J_x^{\rm
ss-F}/E$.

It should be noted that, in our study, the diagonal part of the
distribution function, ${\hat{\cal R}}_0^s({\bf p})$, which is
involved in all disorder-related components of anomalous Hall
current, was evaluated in the self-consistent Born approximation.
This implies that our results correspond to that obtained in the
Kubo formalism by considering the "ladder-sum" vertex corrections to
the bubble diagrams.

\section{Results and discussions}
We have carried out a numerical calculation to investigate the
anomalous Hall effect in a InSb/AlInSb quantum well with Rashba SO
coupling. Such a system was recently examined
experimentally.\cite{Kh} It is well known that the InSb
semiconductor is a good material for AHE observation because its
band gap, $E_0=0.235$\,eV, spin-orbit splitting, $\Delta_{\rm
SO}=0.81$\,eV, and $P=9.63$\,eV$\cdot$\AA \,\,result in a pronounced
spin-orbit coupling constant $\lambda=5.31$\,nm$^{2}$ (for GaAs,
$\lambda=0.053$\,nm$^{2}$).\cite{Parameter} Also, the large
g-factor, $g=-51.4$, may lead to a remarkably large magnetization.
In our calculation, the static dielectric constant, $\kappa$, and
the effective mass of InSb, $m^*$, are chosen to be $\kappa=17.54$
and $m^*=0.0135m_0$ with $m_0$ as the free electron mass. The width
of the InSb/AlInSb quantum well is assumed to be $a=20$\,nm and the
density of electrons is taken as $N_e=1\times 10^{15}$\,m$^{-2}$. We
consider an {\it attractive} interaction between the electrons and
the background impurities in the quantum wells (the attractive and
repulsive interactions lead to differing anomalous Hall effects
because their contributions to AHC in the second Born approximation
have opposite signs). Note that we have also estimated the effect of
scattering of electrons by remote impurities on AHE, finding that it
is relatively small and can be ignored. Thus, the scattering
potential $V_{\bf q}$ can be written as\cite{Lei}
\begin{equation}
V_{\bf q}=U(q)F(q)/\kappa(q,0),
\end{equation}
with $U(q)=-e^2/(2\varepsilon_0\kappa q)$ and the form factor
$F(q)$ determined by ($u=qa$)
\begin{equation}
F(q)=\frac{8\pi^2}{(4\pi^2+u^2)u}\left
[1+\frac{u^2}{4\pi^2}-\frac{1-\exp(-u)}{u}\right].
\end{equation}
$\kappa(q,0)$ is a static dielectric function in random phase
approximation and can be written as
\begin{equation}
\kappa(q,0)=1+\frac{q_s}{q}H(q),
\end{equation}
with $q_s=m^*e^2/(2\pi \epsilon_0\kappa)$ and $H(q)$ given
by\cite{Price}
\begin{equation}
H(q)=3\frac{1-\exp(-u)}{u^2+4\pi^2}+\frac{u}{u^2+4\pi^2}
-\frac{1-\exp(-u)}{(u^2+4\pi^2)^2}(u^2-4\pi^2)
+\frac{2}{u}\left[1-\frac{1-\exp(-u)}{u}\right].
\end{equation}
Here, the effect of the Rashba SO coupling on the screening of
$V_{\bf q}$ is ignored. Further, to determine the impurity
density, we assume that for $M=0$ and $\alpha=0$ the
electron-impurity scattering results in an electron mobility
$\mu_0=5$\,m$^{2}$/Vs.

\subsection{Anomalous Hall effect in a InSb/AlInSb quantum well without Rashba SO coupling}
\begin{figure}
\includegraphics [width=0.45\textwidth,clip] {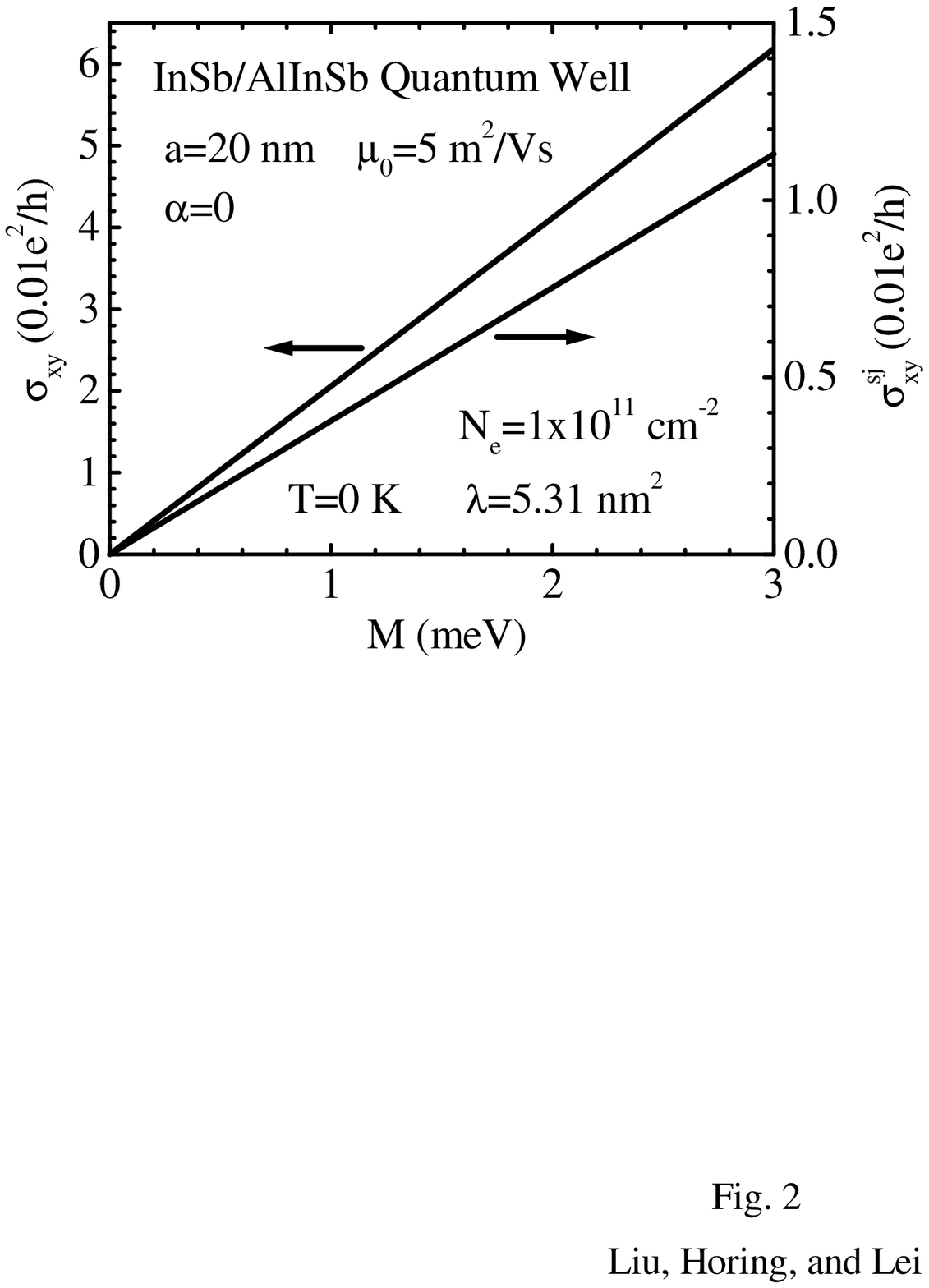}
\caption{Magnetization dependencies of $\sigma_{xy}$ and
$\sigma_{xy}^{\rm sj}$ in a InSb/AlInSb quantum well without
Rashba spin-orbit coupling. The width of the quantum well $a$ is
$a=20$\,nm. The electron density is $N_e=1\times
10^{15}$\,m$^{-2}$. The lattice temperature is $T=0$\,K and the
mobility in the absence of magnetization is $\mu_0=5$\,m$^2$/Vs. }
\label{fig2}
\end{figure}

We first analyzed the anomalous Hall effect in the absence of Rashba
SO interaction. In this case, the lowest order component of
$J_x^{\rm fd}$ in $\lambda$, $J_x^{\rm ss-L}$, vanishes, while the
component $J_x^{\rm ss-F}$ is nonvanishing and reduces to
\begin{equation}
J_x^{\rm ss-F}=e^2E\sum_{{\bf p},\mu}\frac{p}{m^*}\cos^2\phi_{\bf
p}\Phi_{1\mu}^c(p),
\end{equation}
with $\Phi_{1\mu}^c(p)$ determined by
\begin{eqnarray}
\Phi_{1\mu}^c(p)&=&(-1)^{\mu}4\pi^2n_i\tau_{1\mu\mu}\lambda\Phi_{0\mu}^s(p)\sum_{{\bf
k},{\bf q}}V_{{\bf p}-{\bf k}}V_{{\bf k}-{\bf q}}V_{{\bf q}-{\bf
p}}pk\sin(\phi_{\bf k}-\phi_{\bf p})\delta(\varepsilon_{\mu
p}-\varepsilon_{\mu k})\delta(\varepsilon_{\mu p}-\varepsilon_{\mu
q})\nonumber\\
&&\times [pk\sin(\phi_{\bf k}-\phi_{\bf p})+qp\sin(\phi_{\bf
p}-\phi_{\bf q})-qk\sin(\phi_{\bf k}-\phi_{\bf q})],\label{AHEa0ss}
\end{eqnarray}
and $\Phi_{0\mu}^s(p)=-\tau_{1\mu\mu}\frac{\partial
n_F[\varepsilon_\mu(p)]}{\partial
\varepsilon_\mu(p)}\frac{\partial \varepsilon_\mu(p)}{\partial
p}$. Since the contributions to AHC from the off-diagonal elements
of the distribution function, $J_x^{\rm fo}=\left. J_x^{\rm
fo}\right |^I+\left. J_x^{\rm fo}\right |^{II}$, vanish, the
side-jump AHC involves only the components $J_x^{\rm imp}$ and
$J_x^{E}$: $J_x^{\rm sj}=J_x^{\rm imp}+J_x^{E}$. Here, $J_x^{\rm
imp}$ and $J_x^E$ are equal to each other and take the form
\begin{equation}
J_x^{\rm imp}= J_x^{E}={\lambda e^2E}\sum_{\bf p} \left
\{n_F[\varepsilon_1(p)]-n_F[\varepsilon_2(p)]\right\}.\label{AHEa0sj}
\end{equation}

In Fig.\,2, we plot the calculated total anomalous Hall conductivity
$\sigma_{xy}=\sigma_{xy}^{\rm ss}+\sigma_{xy}^{\rm sj}$, and its
component $\sigma_{xy}^{\rm sj}=\sigma_{xy}^{\rm
imp}+\sigma_{xy}^{E}$ as functions of magnetization, $M$. With
increasing magnetization, $\sigma_{xy}$ and $\sigma_{xy}^{\rm sj}$
increase linearly. A comparison between $\sigma_{xy}$ and
$\sigma_{xy}^{\rm sj}$ indicates that, for the given
$\mu_0=5$\,m$^2$/Vs, both the contributions from side jump and skew
scattering are of the same order of magnitude. Note that,
notwithstanding the large spin-orbit coupling constant $\lambda$,
the anomalous Hall conductivity is still much smaller than the
ordinary one: the ordinary Hall conductivity is $34.6$\,$e^2/h$ for
a magnetic field $B=0.34$\,T (in the InSb/AlInSb quantum well with
$g=-51.4$, this magnetic field corresponds to a magnetization
$M=1$\,meV).

From Eqs.\,(\ref{AHEa0ss}) and (\ref{AHEa0sj}) we see that in the
absence of both the Rashba SO coupling and magnetization, the
contributions to anomalous Hall current from electrons with opposite
spins (or helicities) have opposite signs. As a result, the total
anomalous {\it charge} Hall current vanishes. However, there is a
nonvanishing {\it spin} Hall current since electrons with opposite
spins move toward opposite sides of the sample. We estimate the
spin-Hall current in the studied InSb/AlInSb quantum well for
$\alpha=0$ and $M=0$, finding that the spin-Hall mobility $\mu_{sH}$
defined in Ref.\,\onlinecite{Liu2} is $\mu_{sH}=0.013$\,m$^2$/Vs.
(In contrast to this, in a GaAs/AlGaAs quantum well with charge
mobility $\mu_0=0.6$\,m$^2$/Vs, the total spin-Hall mobility is
$\mu_{sH}=-2.0\times 10^{-5}$\,m$^2$/Vs, and the contributions from
side-jump and skew scattering, respectively, are
$\mu_{sH}^{sj}=-1.6\times 10^{-4}$\,m$^2$/Vs and
$\mu_{sH}^{ss}=1.4\times 10^{-4}$\,m$^2$/Vs. They are of the same
order of magnitude as the spin-Hall mobilities in bulk $n$-doped
GaAs: in a bulk GaAs with the same $\mu_0$, Engel {\it et
al.}\cite{Halperin} found $\mu_{sH}^{sj}=-1.6\times
10^{-4}$\,m$^2$/Vs and $\mu_{sH}^{ss}=3.5\times 10^{-4}$\,m$^2$/Vs.)

\subsection{Anomalous Hall effect in a Rashba InSb/AlInSb quantum
well}
We have also calculated the anomalous Hall conductivity in a
InSb/AlInSb quantum well with Rashba spin-orbit interaction. In the
case of nonvanishing $\alpha$, one has to consider not only
$\sigma_{xy}^{\rm ss-F}$ and $\sigma_{xy}^{{\rm imp},E}$, but also
the anomalous Hall conductivities, $\sigma_{xy}^{\rm fo}=\left
.\sigma_{xy}^{\rm fo}\right |^I+\left .\sigma_{xy}^{\rm fo}\right
|^{II}$, and $\sigma_{xy}^{\rm ss-L}$. The results are plotted in
Figs. 3 and 4.

\begin{figure}
\includegraphics [width=0.6\textwidth,clip] {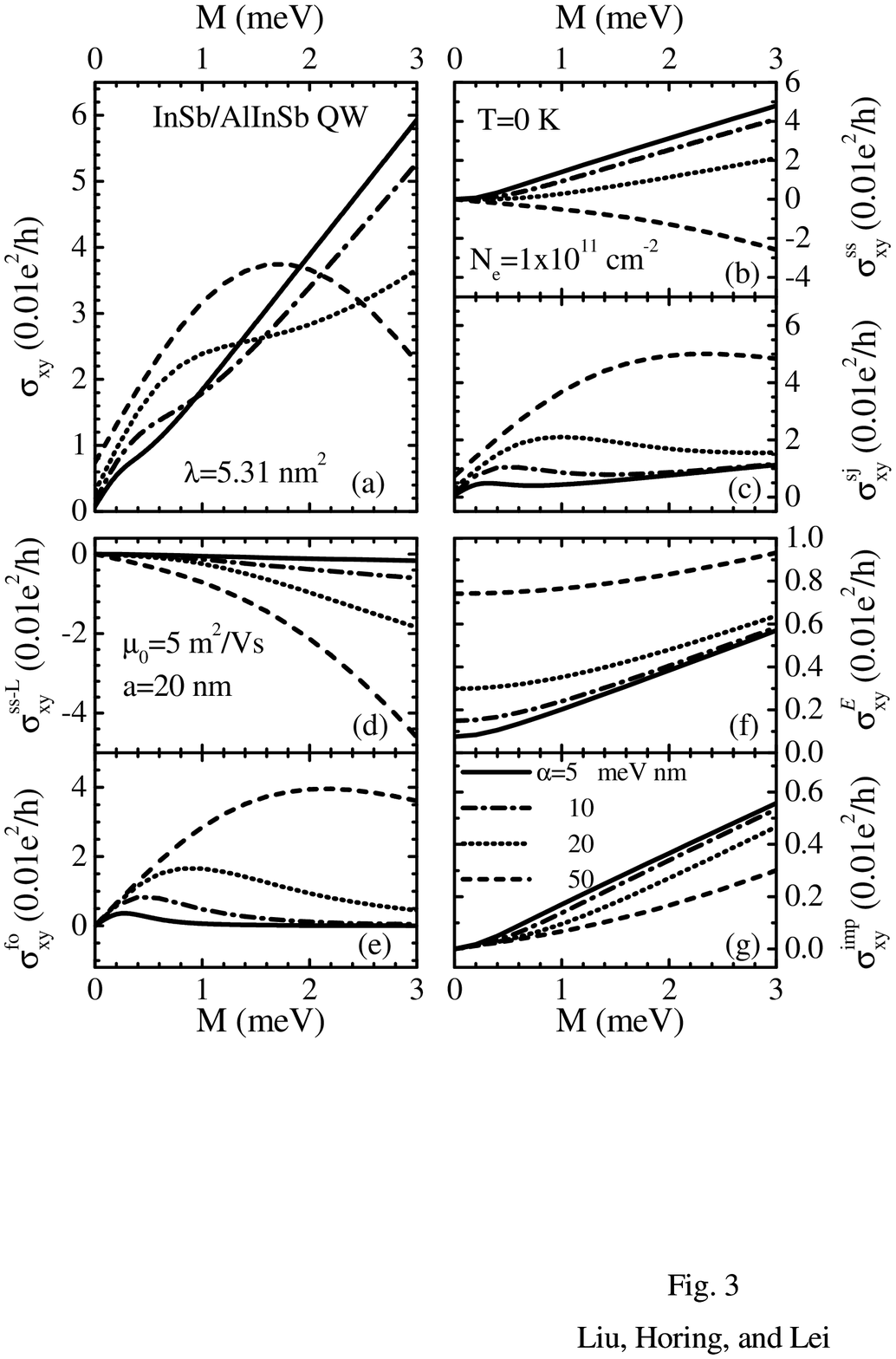}
\caption{Magnetization dependencies of (a) the total anomalous
Hall conductivity, $\sigma_{xy}$, (b) the skew-scattering and (c)
side-jump Hall conductivities, as well as their components, (d)
$\sigma_{xy}^{\rm ss-L} $, (e) $\sigma_{xy}^{\rm fo}$, (f)
$\sigma_{xy}^{E}$ , and (g) $\sigma_{xy}^{\rm imp}$ in a
InSb/AlInSb quantum well with different Rashba SO couplings:
$\alpha=5$, $10$, $20$, and $50$\,meV$\cdot$nm. The other
parameters are the same as in Fig.\,2.} \label{fig3}
\end{figure}
\begin{figure}
\includegraphics [width=0.45\textwidth,clip] {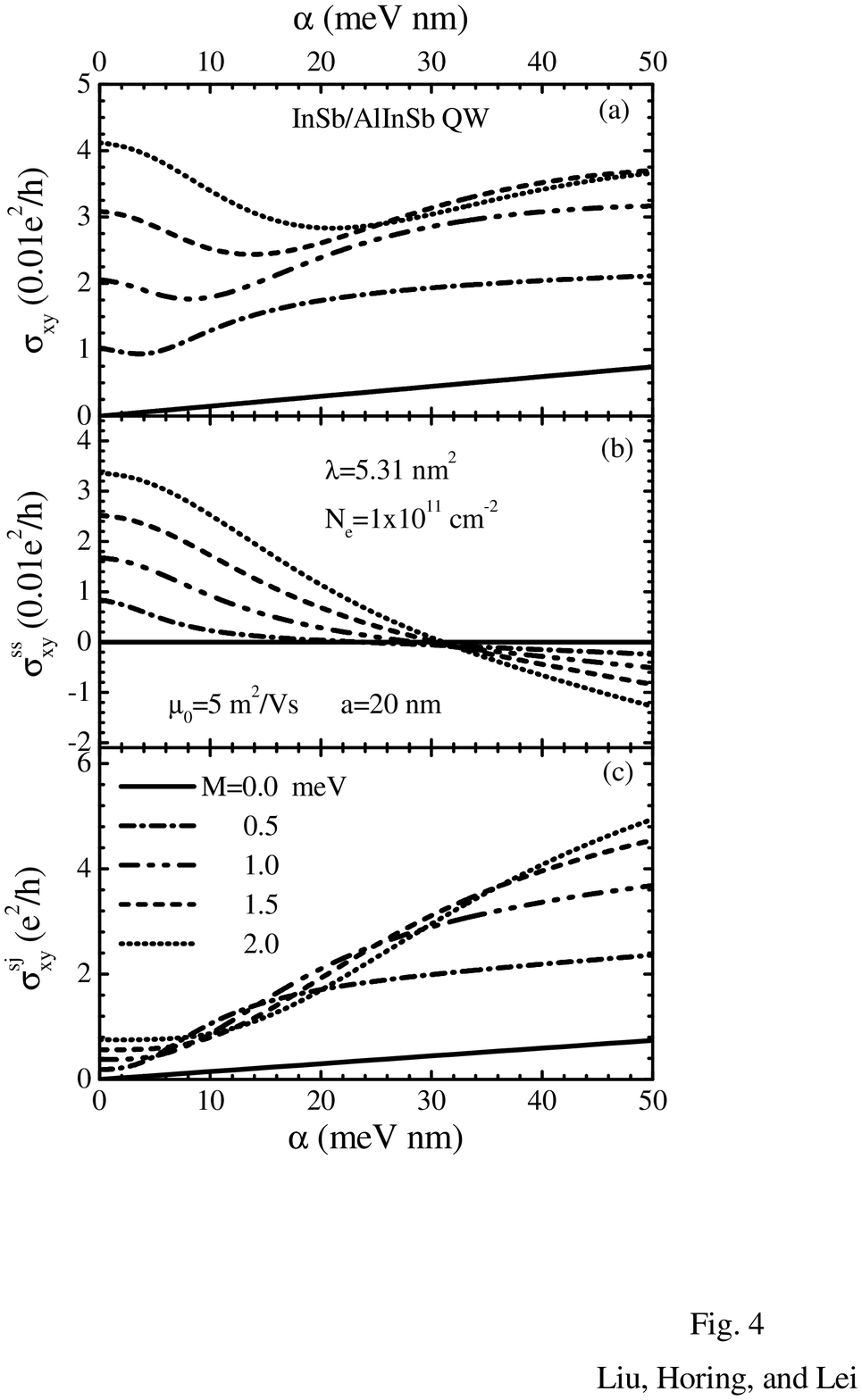}
\caption{The dependencies of (a) the total anomalous Hall
conductivity $\sigma_{xy}$, and  (b) the skew-scattering and (c)
the side-jump anomalous Hall conductivities, $\sigma_{xy}^{\rm
ss}$ and $\sigma_{xy}^{\rm sj}$, on the Rashba spin-orbit coupling
constant for different magnetizations: $M=0$, $0.5$, $1.0$, $1.5$,
and $2.0$\,meV. The other parameters are the same as in Fig.\,2.}
\label{fig4}
\end{figure}
In Fig.\,3, we plot the total anomalous Hall conductivity,
$\sigma_{xy}$, and the skew scattering and side-jump Hall
conductivities, $\sigma_{xy}^{\rm ss}$ and $\sigma_{xy}^{\rm sj}$,
as well as their components, $\sigma_{xy}^{\rm ss-L} $,
$\sigma_{xy}^{\rm fo}$, $\sigma_{xy}^{E}$, and $\sigma_{xy}^{\rm
imp}$, as functions of magnetization $M$ for various Rashba
spin-orbit coupling constants. We find that, as magnetization
increases, the total anomalous Hall conductivity $\sigma_{xy}$
increases for $\alpha=5$, 10, and 20\,meV$\cdot$nm, but it decreases
for $\alpha=50$\,meV$\cdot$nm.

Such complicated behavior of the magnetization dependence of
$\sigma_{xy}$ arises from competition of the side-jump and
skew-scattering contributions to anomalous Hall conductivity. From
Figs.\,3(b) and 3(c) it is obvious that $\sigma_{xy}^{\rm ss}$
varies monotonically with the magnetization, but in the
magnetization dependence of $\sigma_{xy}^{\rm sj}$ there is always a
small dip (when $\alpha=50$\,meV$\cdot$nm, this dip is shifted out
of the studied magnetization range). In $\sigma_{xy}^{\rm ss}$,
$\sigma_{xy}^{\rm ss-F}$ is dominant for small $\alpha$, leading to
an increase of $\sigma_{xy}^{\rm ss}$ with increasing $M$. However,
when $\alpha>30$\,meV$\cdot$nm, $\sigma_{xy}^{\rm ss-L}$ is
important: $\sigma_{xy}^{\rm ss}$ becomes negative and its
magnetization dependence exhibits a decrease with increasing
magnetization. Among the contributions to $\sigma_{xy}^{\rm sj}$,
$\sigma_{xy}^{\rm fo}$ is dominant for large magnetization, while
$\sigma_{xy}^{ E}$ is important for small $M$ and results in a
nonvanishing $\sigma_{xy}^{\rm sj}$ when $\alpha\neq 0$ but $M=0$.

In Fig.\,4, we plot $\sigma_{xy}$, $\sigma_{xy}^{\rm ss}$, and
$\sigma_{xy}^{\rm sj}$ as functions of the spin-orbit coupling
constant. We find that, as $\alpha$ increases, $\sigma_{xy}$
increases monotonically for $M=0$\,meV, while for $M=0.5-2$\,meV,
$\sigma_{xy}$ first decreases and then increases. It is also evident
from Figs.\,4(b) and 4(c) that, with increasing Rashba spin-orbit
coupling constant $\alpha$, $\sigma_{xy}^{\rm sj}$ increases while
$\sigma_{xy}^{\rm ss}$ decreases. Since the rate of increase or
decrease depends on $M$, the $\alpha$ dependence of the total
anomalous Hall conductivity behaves differently for different
magnetizations.

\section{Conclusions}
We have employed a kinetic equation approach to investigate the
anomalous Hall effect in Rashba 2D electron systems based on narrow
band semiconductors. The Rashba SO coupling was considered
nonperturbatively, while the extrinsic spin-orbit interaction and
the SO coupling directly induced by an external driving electric
field were taken into account in the first order of the coupling
constant. Considering electron-impurity scattering up to the
second-Born approximation, we found that the various components of
the anomalous Hall current can fit into two classes: the side-jump
and skew scattering anomalous Hall currents. The side-jump anomalous
Hall current involves contributions not only from the extrinsic SO
coupling, but also from SO coupling directly induced by the driving
electric field. It also contains a component which arises from
Rashba SO coupling and relates to the off-diagonal elements of the
helicity-basis distribution function. The skew-scattering AHE arises
from the anisotropy of the diagonal elements of the distribution
function, and it is a result of the Rashba and extrinsic SO
interactions. We also performed a numerical calculation to
investigate the anomalous Hall effect in a InSb/AlInSb quantum well.
We found that the contributions to anomalous Hall conductivity from
both the side-jump and skew scattering terms are of the same order
of magnitude, leading to complicated dependencies of the total
anomalous Hall conductivity on magnetization and on the Rashba
spin-orbit coupling constant. It is also clear that the component
arising from the SO coupling due to the driving electric field is
dominant for small magnetization.

\begin{acknowledgments}
This work was supported by the Department of Defense through the
DURINT program administered by the US Army Research Office, DAAD
Grant No. 19-01-1-0592, and by projects of the National Science
Foundation of China and the Shanghai Municipal Commission of
Science and Technology.
\end{acknowledgments}

\appendix*
\section{Explicit form of $\hat I^{(1)}$}

The diagonal elements of $\hat I^{(1)}$, $(\hat
I^{(1)})_{\mu\mu}$, can be written as sums of four terms: $(\hat
I^{(1)})_{\mu\mu}= {\cal I}^a_{\mu}[\hat \rho^I_1]+{\cal
I}^{b}_{\mu}[\hat \rho^I_1]+{\cal I}^{c}_{\mu}[\hat
\rho^I_1]+{\cal I}^{d}_{\mu}[\hat \rho^I_1]$ ($[\hat \rho^I_1]$
denotes that ${\cal I}^{a,b,c,d}_{\mu}$ depend on the specific
form of the distribution function $\hat \rho^I_1$). ${\cal
I}^a_{\mu}[\hat \rho^I_1]$ and ${\cal I}^{b}_{\mu}[\hat \rho^I_1]$
are the terms of the first Born approximation and can be written
as
\begin{eqnarray}
{\cal I}^a_\mu[\hat \rho^I_1]&=&\pi n_i\sum_{\bf k} |V({\bf p}
-{\bf k})|^2 \left \{a_+({\bf p},{\bf k})
[(\rho^{I}_1)_{\mu\mu}({\bf p})-(\rho^{I}_1)_{\mu\mu}({\bf k})]
\Delta_{\mu\mu}
\right.\nonumber\\
&&\left . + a_-({\bf p},{\bf k})[(\rho^{I}_1)_{\mu\mu}({\bf
p})-(\rho^{I}_1)_{{\bar \mu}{\bar \mu}}({\bf k})]\Delta_{\mu{\bar
\mu}}\right \},\label{DDFs}
\end{eqnarray}
and
\begin{eqnarray}
{\cal I}^{b}_\mu[\hat \rho^I_1]&=&{2\pi n_i\lambda\alpha^2}
\sum_{\bf k} |V({\bf p} -{\bf k})|^2
\frac{k^2p^2}{\lambda_p\lambda_k}\sin^2(\phi_{\bf k}-\phi_{\bf
p})\nonumber\\
&&\times\left \{[(\rho^{I}_1)_{\mu\mu}({\bf
p})-(\rho^{I}_1)_{\mu\mu}({\bf k})]
\Delta_{\mu\mu}-[(\rho^{I}_1)_{\mu\mu}({\bf
p})-(\rho_1^{I})_{{\bar \mu}{\bar \mu}}({\bf k})]\Delta_{\mu{\bar
\mu}}\right \},\label{DDFsii}
\end{eqnarray}
with $\Delta_{\mu\nu}\equiv \delta[\varepsilon_\mu({\bf
p})-\varepsilon_\nu({\bf k})]$ and  $a_\pm({\bf p},{\bf k})\equiv
(\lambda_p\lambda_k\pm M^2\pm \alpha^2kp\cos (\phi_{\bf p}-\phi_{\bf
k}))/ \lambda_p\lambda_k $. ${\cal I}^{c}_\mu[\hat \rho^I_1]$ and
${\cal I}^{d}_\mu[\hat \rho^I_1]$ are the terms of the second Born
approximation. ${\cal I}^{c}_\mu[\hat \rho^I_1]$ is explicitly
independent of the spin-orbit coupling constant $\lambda$ and takes
the form
\begin{eqnarray}
{\cal I}^{c}_\mu[\hat \rho^I_1]&=&\pi^2 Mn_i\alpha^2\sum_{{\bf
k},{\bf q}}V_{{\bf p} -{\bf k}}V_{{\bf k} -{\bf q}}V_{{\bf q}
-{\bf
p}}\frac{1}{\lambda_p\lambda_k}\nonumber\\
&&\times \left \{c_{\mu +}({\bf p},{\bf q},{\bf
k})\Delta_{\mu\mu}(\rho^{I}_1)_{\mu\mu}({\bf k})+c_{\mu -}({\bf
p},{\bf q},{\bf k})\Delta_{\mu\bar \mu}(\rho^{I}_1)_{\bar \mu\bar
\mu}({\bf k})\right \},\label{Ic}
\end{eqnarray}
while ${\cal I}^{d}_\mu[\hat \rho^I_1]$ is linear in $\lambda$ and
can be written as
\begin{eqnarray}
{\cal I}^{d}_\mu[\hat \rho^I_1]&=&\pi^2 \lambda Mn_i\sum_{{\bf
k},{\bf q}}V_{{\bf p} -{\bf k}}V_{{\bf k} -{\bf q}}V_{{\bf q} -{\bf
p}}\frac{1}{\lambda_p\lambda_k}\nonumber\\
&&\times \left \{d_{\mu +}({\bf p},{\bf q},{\bf
k})\Delta_{\mu\mu}(\rho^{I}_1)_{\mu\mu}({\bf k})+d_{\mu -}({\bf
p},{\bf q},{\bf k})\Delta_{\mu\bar \mu}(\rho^{I}_1)_{\bar \mu\bar
\mu}({\bf k})\right \}.
\end{eqnarray}
The parameters $c_{\mu \pm}({\bf p},{\bf k},{\bf q})$ and $d_{\mu
\pm}({\bf p},{\bf k},{\bf q})$ are defined as
\begin{eqnarray}
c_{\mu \pm}({\bf p},{\bf q},{\bf
k})&=&\mp\frac{(-1)^\mu}{\lambda_q}{{\cal C}}({\bf p},{\bf q},{\bf
k})[\delta(\varepsilon_\mu({\bf p})-\varepsilon_\mu({\bf
q}))-\delta(\varepsilon_\mu({\bf p})-\varepsilon_{\bar \mu}({\bf
q}))],\label{c}
\end{eqnarray}
\begin{eqnarray}
d_{\mu \pm}({\bf p},{\bf k},{\bf
q})&=&\frac{1}{\lambda_q}\left\{{{\cal D}}^m_{\mu\pm}({\bf p},{\bf
q},{\bf k})[\delta(\varepsilon_\mu({\bf p})-\varepsilon_\mu({\bf
q}))-\delta(\varepsilon_\mu({\bf p})-\varepsilon_{\bar \mu}({\bf
q}))]\right.\nonumber\\
&&\left .+{{\cal D}}^p_{\mu\pm}({\bf p},{\bf q},{\bf
k})[\delta(\varepsilon_\mu({\bf p})-\varepsilon_\mu({\bf
q}))+\delta(\varepsilon_\mu({\bf p})-\varepsilon_{\bar \mu}({\bf
q}))]\right\},\label{d}
\end{eqnarray}
with
\begin{equation}
{\cal C}({\bf p},{\bf q},{\bf k})=pq\sin(\phi_{\bf p}-\phi_{\bf
q})-qk\sin(\phi_{\bf k}-\phi_{\bf q})+pk\sin(\phi_{\bf
k}-\phi_{\bf p}),
\end{equation}
\begin{eqnarray}
{\cal D}^m_{\mu\pm}({\bf p},{\bf q},{\bf k})&=&(-1)^{\mu+1}
(\pm M^2+\lambda_k\lambda_p)[pk\sin(\phi_{\bf k}-\phi_{\bf
p})+qp\sin(\phi_{\bf p}-\phi_{\bf
q})-qk\sin(\phi_{\bf k}-\phi_{\bf
q})]\nonumber\\
&&\pm (-1)^{\mu+1}\alpha^2\left \{pk\sin(\phi_{\bf k}-\phi_{\bf
p})[pq\cos(\phi_{\bf p}-\phi_{\bf q})+qk\cos(\phi_{\bf
q}-\phi_{\bf k})-kp\cos(\phi_{\bf
k}-\phi_{\bf p})]\right.\nonumber\\
&&+qp\sin(\phi_{\bf p}-\phi_{\bf
q})[-pq\cos(\phi_{\bf p}-\phi_{\bf q})+qk\cos(\phi_{\bf
q}-\phi_{\bf k})+kp\cos(\phi_{\bf
k}-\phi_{\bf p})]
\nonumber\\
&&\left. +qk\sin(\phi_{\bf k}-\phi_{\bf q})[-pq\cos(\phi_{\bf
p}-\phi_{\bf q})+qk\cos(\phi_{\bf q}-\phi_{\bf
k})-kp\cos(\phi_{\bf k}-\phi_{\bf p})] \right\},\nonumber\\
\end{eqnarray}
and
\begin{equation}
{\cal D}^p_{\mu\pm}({\bf p},{\bf q},{\bf k})=(-1)^{\mu+1}
\lambda_q(\lambda_k\pm\lambda_p)[pk\sin(\phi_{\bf k}-\phi_{\bf
p})+qp\sin(\phi_{\bf p}-\phi_{\bf q})-qk\sin(\phi_{\bf
k}-\phi_{\bf q})].
\end{equation}

Since the off-diagonal elements of the collision term $\hat I^{(1)}$
are simply related by $\hat I^{(1)}_{12}=-[\hat I^{(1)}_{12}]^*$, it
suffices to consider the element $\hat I^{(1)}_{12}$. In the
first-Born approximation, $\hat I^{(1)}_{12}$ can be expressed as a
sum of two terms: $\hat I^{(1)}_{12}={\cal S}^{a}[\hat
\rho^I_1]+{\cal S}^{b}[\hat \rho^I_1]$ with ${\cal S}^{a}[\hat
\rho^I_1]$ and ${\cal S}^{b}[\hat \rho^I_1]$ as the terms in the
lowest- and first-order of $\lambda$, respectively, and determined
by
\begin{eqnarray}
{\cal S}^{a,b}[\hat \rho^I_1]&=&\frac{n_i}{2}\sum_{{\bf k}}|V({\bf
p}-{\bf k})|^2g_{a,b}({\bf p},{\bf k})\left \{ \left [(\hat {\rm
G}^r_{0k})_{11}(\varepsilon_{2 p}) -(\hat {\rm
G}^r_{0k})_{22}(\varepsilon_{2 p})\right ]
(\hat \rho_1^I)_{22}({\bf p})\right.\nonumber\\
&&\left .-\left [(\hat {\rm G}^a_{0p})_{22}(\varepsilon_{2 k})
-(\hat {\rm G}^r_{0p})_{11}(\varepsilon_{2 k})\right ] (\hat
\rho_1^I)_{22}({\bf k})-\left [(\hat {\rm
G}^a_{0k})_{11}(\varepsilon_{1 p}) -(\hat {\rm
G}^a_{0k})_{22}(\varepsilon_{1 p})\right ] (\hat
\rho_1^I)_{11}({\bf p})\right.\nonumber\\
&&\left .+\left [(\hat {\rm G}^a_{0p})_{22}(\varepsilon_{1 k})
-(\hat {\rm G}^r_{0p})_{11}(\varepsilon_{1 k})\right ] (\hat
\rho_1^I)_{11}({\bf k}) \right\}.\label{NDF}
\end{eqnarray}
Here, $g_{a}({\bf p},{\bf k})\equiv [\alpha k\lambda_{p}\sin
(\phi_{\bf p}-\phi_{\bf k}) +i\alpha M(p- k \cos(\phi_{\bf
p}-\phi_{\bf k}))]/(\lambda_k\lambda_p)$ and $g_{b}({\bf p},{\bf
k})\equiv -2\sin(\phi_{\bf p}-\phi_{\bf k})[iM\sin (\phi_{\bf
p}-\phi_{\bf k}) +\lambda_p \cos(\phi_{\bf p}-\phi_{\bf
k}))]\lambda\alpha k^2 p/(\lambda_k\lambda_p)$.

\end{document}